\documentclass[8pt,twocolumn]{extarticle}

\usepackage{authblk}
\usepackage{graphicx}
\usepackage{float}
\usepackage{amsmath}
\usepackage{xcolor}
\usepackage{booktabs}
\usepackage[square,numbers,sort&compress]{natbib}
\usepackage{enumitem}
\usepackage{hyperref}

\usepackage{relsize}
\def\eg{{\it{e.g.,~}}}
\def\ie{{\it{i.e.,}}}
\def\HAR{\textsc{HAR}}
\def\FIF{\textsc{FIF}}
\def\RAN{\textsc{RAN}}
\def\TRANS{\textsc{TRANS}}
\def\MIN{\textsc{MIN}}
\def\ANY{\textsc{ANY}}
\def\DAT{\textsc{DAT}}
\def\Imin{I_\textrm{min}}
\def\IP{I_\textrm{A}}

\def\fD{f_\textrm{D}}

\title{Cross-cultural data shows musical scales evolved to maximise imperfect fifths}

\author[1,*]{John M. McBride}
\author[1,2,*]{Tsvi Tlusty}
\affil[1]{Center for Soft and Living Matter, Institute for Basic Science, Ulsan 44919, South Korea}
\affil[2]{Department of Physics, Ulsan National Institute of Science and Technology, Ulsan 44919, South Korea}
\affil[*]{jmmcbride@protonmail.com, tsvitlusty@gmail.com}

\begin{document}

\maketitle


\textbf{
  Musical scales are used throughout the world,
  but the question of how they evolved remains open. 
  Some suggest that scales based on the harmonic series are
  inherently pleasant, while others propose that scales
  are chosen that are easy to communicate. 
  However, testing these theories has been hindered by 
  the sparseness of empirical evidence.
  Here, we assimilate data from diverse
  ethnomusicological sources into a cross-cultural database of scales.
  We generate populations of scales based on multiple theories
  and assess their similarity to empirical distributions from the database.
  Most scales tend to include intervals which are
  close in size to perfect fifths (``imperfect fifths''), 
  and packing arguments explain the salient features of the distributions.
  Scales are also preferred if their intervals are compressible,
  which may facilitate efficient communication and memory of melodies.
  While scales appear to evolve according to various selection pressures,
  the simplest imperfect-fifths packing model best fits the empirical data.
}

  How and why did humans evolve to create and appreciate music? 
  The question arose at the dawn of evolutionary theory and been asked ever since
  \cite{dar96, mcdmp05, fitco06, bal10}.
  Studying musical features that are conserved across cultures may lead to possible answers.
  \cite{wal01, net10}.
  Two such universal features are the use of discrete pitches and the octave, 
  defined as an interval between two notes 
  where one note is double the frequency of the other \cite{bropm13}. 
  Taken together, these form the musical scale, defined as a set of intervals
  spanning an octave (Fig. \ref{fig:model}A). 
  Musical scales can therefore be considered solutions to the problem of
  partitioning an octave into intervals, and thus can be treated mathematically.
  Examination of scales from different cultures can help elucidate the basic
  perception and production mechanisms that humans share and shed light on this
  evolutionary puzzle.

  One theory on the origin of scales suggests that the frequency
  ratios of intervals in a scale ought to consist of simple integers \cite{gilpo09}.
  After the octave (2:1), the simplest ratio is 3:2, referred to
  in Western musical theory as a perfect fifth.
  Frequencies related by simple integer ratios naturally occur
  in the harmonic series -- a plucked string will produce a complex sound with
  a fundamental frequency accompanied by integer multiples of the fundamental.
  The theory that scales are related to harmonicity follows from the idea that
  exposure to harmonic sounds in animal vocalisations
  may have conditioned humans to respond positively to them \cite{schjn03,brucb10}.
  Musical features related to harmonicity -- harmonic intervals \cite{wesml94,rec18,jiepf19},
  octave equivalence \cite{bur99,pat10}, a link between consonance and
  harmonicity  \cite{mcdcb10,coupn12,bowpn15} -- are indeed widespread,
  although their universality is disputed \cite{mcdna16,jaccb19}.
  One study attempted to explain the origin of scales as maximization
  of harmonicity \cite{gilpo09}, however the universality of their findings
  is limited by the scope of cultures considered \cite{sav19}.

  The vocal mistuning theory states that scales, and intervals, were
  chosen not due to harmonicity, but because they were easy to
  communicate \cite{pfojc17,satfm19}.
  We perceive intervals as categories \cite{siepp77,burja78,lynje91,mcdja10},
  and due to errors in producing \cite{seris11,devpm11,pfojc17},
  and perceiving notes \cite{ruspp05,ruspb05,arump14,gerjr15,larbr19},
  musical intervals are not exact frequency ratios, but rather they
  span a range of acceptable interval sizes. Any overlap between interval categories
  will then result in errors in transmission. This theory appears promising,
  but it has not yet been rigorously investigated.

  These two theories were proposed separately, yet
  they are not mutually exclusive. In this paper, we modify, integrate and expand
  upon these ideas to construct a general, stochastic model which generates populations
  of scales. Our aim is to test which model best mimics scales created by humans.
  To this end, we assembled the most diverse and extensive database of scales from 
  ethnomusicological records. By comparing model-generated
  theoretical distributions with the empirical distributions, we find
  that the theory that best fits the data is the simplest.
  Most scales are arranged to maximise inclusion of imperfect fifths -- perfect fifths
  with a tolerance for error.
  Scales are often found to be compressible, which may make them easier to transmit.
  Adding more detail, beyond fifths, to harmonicity-based theories decreased their performance,
  which suggests that only the first few harmonics are significant in this context.

\section*{Results}

\subsubsection*{Harmonicity Models}

  The main assumption underpinning the harmonicity theory is that human pitch processing
  evolved to take advantage of natural harmonic sounds \cite{ploja64}.
  For example, harmonic amplitudes typically decay with harmonic number
  \cite{schjn03,pat10}, and correspondingly, lower harmonics tend to be
  more dominant in pitch perception \cite{pla05}.
  Thus many proxy measures of harmonicity contain parameters to
  account for harmonic decay \cite{harpr19}. However, 
  to minimize a priori assumptions and model parameters, we avoid explicitly modelling
  harmonic decay. Instead, we test two simple harmonicity theories
  which differ in how they treat higher order harmonics.

\begin{figure} [t!] \centering
  \includegraphics[width=0.85\linewidth]{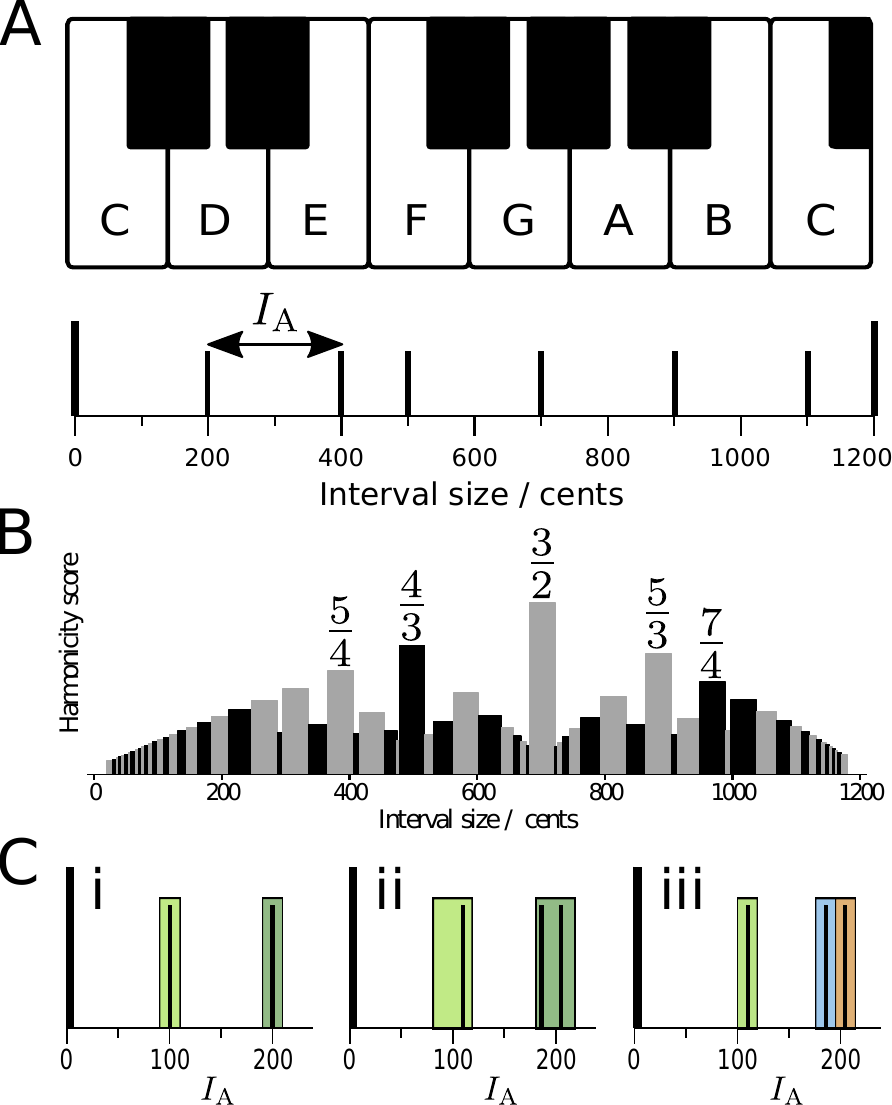}
  \caption{ \label{fig:model}
  A: The major scale starting on a C shown on a piano (equal temperament),
  and as interval sizes (in cents)
  of each note measured from the first note. Intervals between adjacent pairs of notes
  are denoted by $\IP$.
  B: We use a harmonicity template to assign harmonicity scores to intervals \cite{gilpo09}.
  The highest scoring intervals (frequency ratios are shown for the top five)
  act as windows of attraction, whereby any interval in this window is assigned its score.
  In this case the maximum window size is $w=40$ cents.
  C: Adjacent interval, $\IP$, sets 
  for the major scale in equal temperament (i) and just intonation (ii \& iii).
  Boxes indicate interval categories, where the width represents the error,
  and boxes with similar colours are related by a common denominator:
  (i) is losslessly compressed; a similar compression of
  (ii) is lossy; (iii) is uncompressed and therefore costs more to transmit.
  }
  \end{figure}

  The first harmonicity model (\HAR),
  assumes that there is no harmonic decay, for which the model of
  reference \cite{gilpo09} is appropriate.
  This model scores an interval $f_2/f_1$ defined by two frequencies,
  $f_1$ and $f_2$, based on the fraction of harmonics
  of $f_2$ that are matched with the harmonics of $f_1$ in an infinite series.
  Humans do not notice small deviations from simple ratios \cite{harja93,loomp95,carlm94},
  and the model accounts for this by considering intervals
  as categories of width $w$ cents; intervals are measured in cents such that
  an octave is $1200 \log_2 \frac{f_2}{f_1} =1200$ cents. Intervals
  are assigned to a category according to the highest scoring interval within
  $w/2$ cents. The resulting template (Fig. \ref{fig:model}B) is used to calculate
  the average harmonicity score for each scale across all $N\times(N-1)$
  possible intervals, apart from the octave; $N$ is the number of notes in a scale.
  We make no assumptions about tonality, and thus all
  intervals are weighted equally.
  The \HAR~model assumes that scales evolved to maximise this harmonicity score. 

  The second harmonicity model (\FIF) considers the limiting case of high harmonic
  decay. As harmonic decay increases, eventually a few intervals in a harmonic series
  become dominant (SI Table 1), in the order of unison, octave, fifth, etc.
  Thus, this model assumes that due to harmonic decay, \textit{only the octave
  and the fifth} significantly affected the evolution of scales.
  Given this, we simply count the fraction of intervals that are fifths, out
  of all $N\times(N-1)$ possible intervals -- we do not count the octave,
  and we make no assumptions about tonality.
  We allow a tolerance for
  errors, $w$, and thus define ``imperfect fifths'' as intervals
  of size $702 \pm w/2$ cents.
  The \FIF~model assumes that scales evolved to maximise the number of imperfect fifths
  that can be formed in a scale.

\subsubsection*{Transmittability Model}

  The transmittability theory assumes that intervals are perceived as broad categories,
  and the humans make errors in both production and perception of intervals.
  Intervals must thus be large enough to avoid errors in transmission
  due overlapping interval categories (SI Fig. 1).
  This is not sufficient, however, to explain the considerable convergence
  in scales across cultures. We can further consider that scales are optimized
  for minimizing errors in transmission by favouring
  the use of large intervals, however this bias exclusively favours
  equidistant scales (SI pg. 4).
  While equidistant scales do exist \cite{wacna50,vanam80,mcnja08,rosms17},
  they are a minority \cite{savpn15}, so additional mechanisms
  are needed to explain the origin of scales.

  When humans encode continuous audiovisual information such
  as speech, musical rhythm, brightness, or color, there is evidence that it
  is done efficiently \cite{lewnn02,despe03,ravnh16,jaccb17,stipn10,regpn07,badps09}.
  If the same is true for pitch, then compressible scales would
  facilitate communication of melodies.
  An example of a compressible scale is the equal temperament
  major scale (Fig. \ref{fig:model}A).
  We can represent this scale using its notes (C, D, etc.) or as a sequence
  of adjacent intervals, $\IP$: $200, 200, 100, 200, 200, 200, 100$.
  The most compressed representation uses an alphabet size of one
  by encoding the large interval ($200$) in terms of the small one ($100$) (Fig. \ref{fig:model}Ci).
  However for the just intonation tuning -- $\IP$: $204, 182, 112, 204, 182, 204, 112$ -- this code is
  lossy (Fig. \ref{fig:model}Cii). Lossless compression would require a three letter alphabet
  (Fig. \ref{fig:model}Ciii). 
  
  With this in mind, we create a third model (\TRANS) to test whether scales evolved
  to be compressible.
  We define scale compressibility as how accurately an $\IP$ set can be
  represented by a simple interval category template.
  We consider templates with categories centred about integer multiples
  of a common denominator (e.g., $100$, $200$, $300$). Accuracy in this case corresponds to
  the distance between $\IP$ and the centre of the closest category.
  Note that this is merely an approximation of information-theoretic 
  compressibility as we do not explicitly calculate the information content.
  The \TRANS~model assumes that scales are selected for their ease of
  communication, and that this is captured by our measure of compressibility.

\subsubsection*{Generative Monte-Carlo simulations}

  To test the theories, we use Monte Carlo simulations to generate populations of scales,
  based on the above discussion on harmonicity and transmittability.
  We impose a minimum interval size constraint, $\Imin$, such that 
  scales with intervals $<\Imin$ are rejected. Depending on the theory,
  we accept or reject scales according to a cost function and a 
  corresponding Boltzmann probability, which allows us to control
  the strength of the bias via a parameter $\beta$.
  As $\beta$ increases, the generated populations become increasingly
  selective and eventually too selective, resulting in 
  an optimal $\beta$ at which the generated scales best matches the empirical scales.

  We examine and compare results for five models:
  \begin{itemize}[leftmargin=0.50cm]
    \setlength\itemsep{-0.30em}
    \item \RAN: random scales subject to no constraints
                or biases. 
    \item \MIN: random scales with a minimal interval constraint.
    \item \FIF: scales that maximize the number of imperfect fifths.
    \item \HAR: scales biased to have high harmonic similarity score.
    \item \TRANS: transmittable scales biased to be compressible.
  \end{itemize}

  To simplify our simulations, we assume that adjacent
  intervals, $\IP$, in a scale add up to an octave
  of fixed size, $1200$ cents. We fix the number of notes, $N$,
  as a model parameter. In addition, 
  (i) The \FIF, \HAR, and \TRANS~models are also subject to the 
  $\Imin$ constraint of the \MIN~model; 
  (ii) The \FIF~and \HAR~models include the maximum window size $w$
  as a variable parameter; 
  (iii) The \TRANS~model has a single parameter, $n$, 
  that affects the way inaccuracies due to compression are penalized.
  
  For each model we generate a sample size of $S=10^4$ scales. 
  Unless stated otherwise, we show results for $\Imin$ = $80$ cents, $w=20$ and $n=2$.
  We varied the parameters $\Imin$, $w$ and $n$,
  and found that differences are not negligible but do not alter
  the main results of this work (SI Fig. 2). We optimized the strength
  of the bias for each model and each $N$ by tuning $\beta$ (SI Table 4).

  \begin{figure}  \centering
  \includegraphics[width=1.00\linewidth]{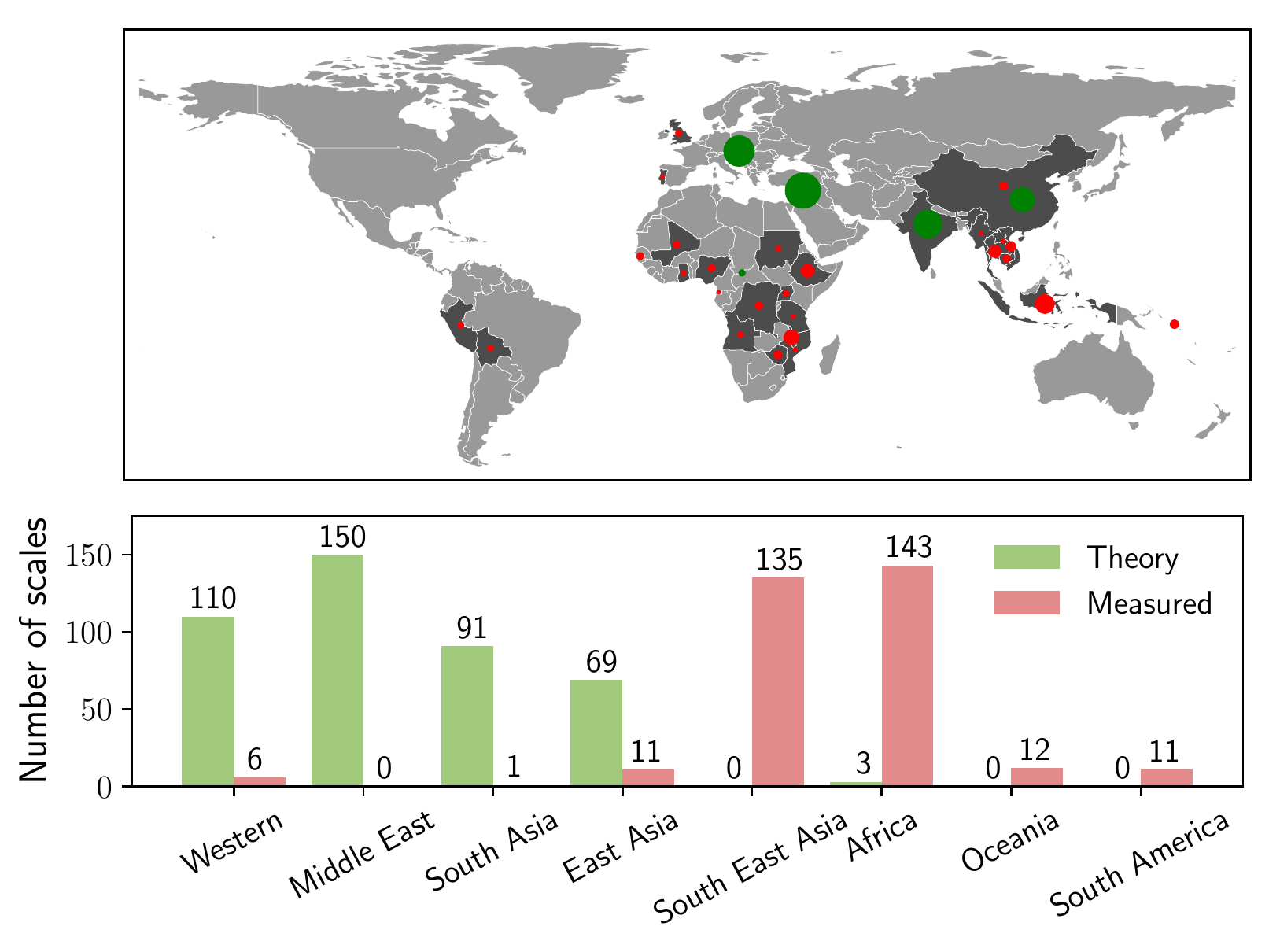}
  \caption{ \label{fig:world}
  Scales in the database either come from a mathematical theory (theory)
  or from measurements from instruments or recordings (measured). The map shows
  the origin of the scales (theory: continents; measured: dark shaded
  countries), with sample size, $S$, indicated
  by the marker size.
  }
  \end{figure}

\subsubsection*{Scale database}
 
  To evaluate the models, we created a database with the aim of recording the diversity
  of scales used by different cultures.
  To this end, we compiled scales that
  use exact mathematical ratios for interval sizes (\eg Western,
  Arabic, Carnatic), taking into account usage of different
  tuning systems (\eg equal temperament, just intonation) \cite{rec18}. 
  Despite the wealth of scales of this type, these cultures are but
  a fraction of the those that produce music.
  In our aim to obtain a comprehensive, diverse database we amalgamated
  work from ethnomusicologists who measured tuning systems used across the world. 
  The database is split into `theory' scales (sample size, $S=423$)
  which have exact theoretical values for frequency ratios,
  and `measured' scales ($S=319$) which were inferred from measurements of instrument
  tunings and recorded performances (Fig. \ref{fig:world}).
  `Measured' scales within a culture can vary significantly
  (e.g., the Gamelan slendro scale),
  and given our goal of capturing this diversity, we recorded
  multiple `measured' scales even if they have the same name.
  A full list of references, inclusion criteria, and
  the numbers/types of scales taken from each reference, are
  presented in the SI, Table 2 and 3. 
  
  A caveat with this approach is that the database
  might reflect hidden biases. For example, imperial dominance
  and globalization can result in homogenization
  of cultures \cite{cow09}, and many ethnomusicologists were biased
  towards reporting on cultures that were more distinct than similar \cite{net10}.
  It is therefore inherently difficult to rigorously define 
  a `correct' empirical distribution of scales, 
  but we believe that this is a suitable approach to start with.
  Another issue is that despite the wide coverage of the database,
  it includes only a fraction of cultures, both geographically and historically, 
  so it can be considered a lower bound on the diversity of scales.
  These limitations may be overcome with the aid of tools that can
  reliably estimate scales from ethnographic recordings, which will
  enable studies on a larger scale \cite{seris11, sixis11}.
  By making our database open we hope to inspire others to plug
  these gaps and undertake further quantitative analysis of musical scales.

\subsubsection*{Packing intervals into octaves partially explains our choices of adjacent intervals}

\begin{figure}  \centering
  \includegraphics[width=1.00\linewidth]{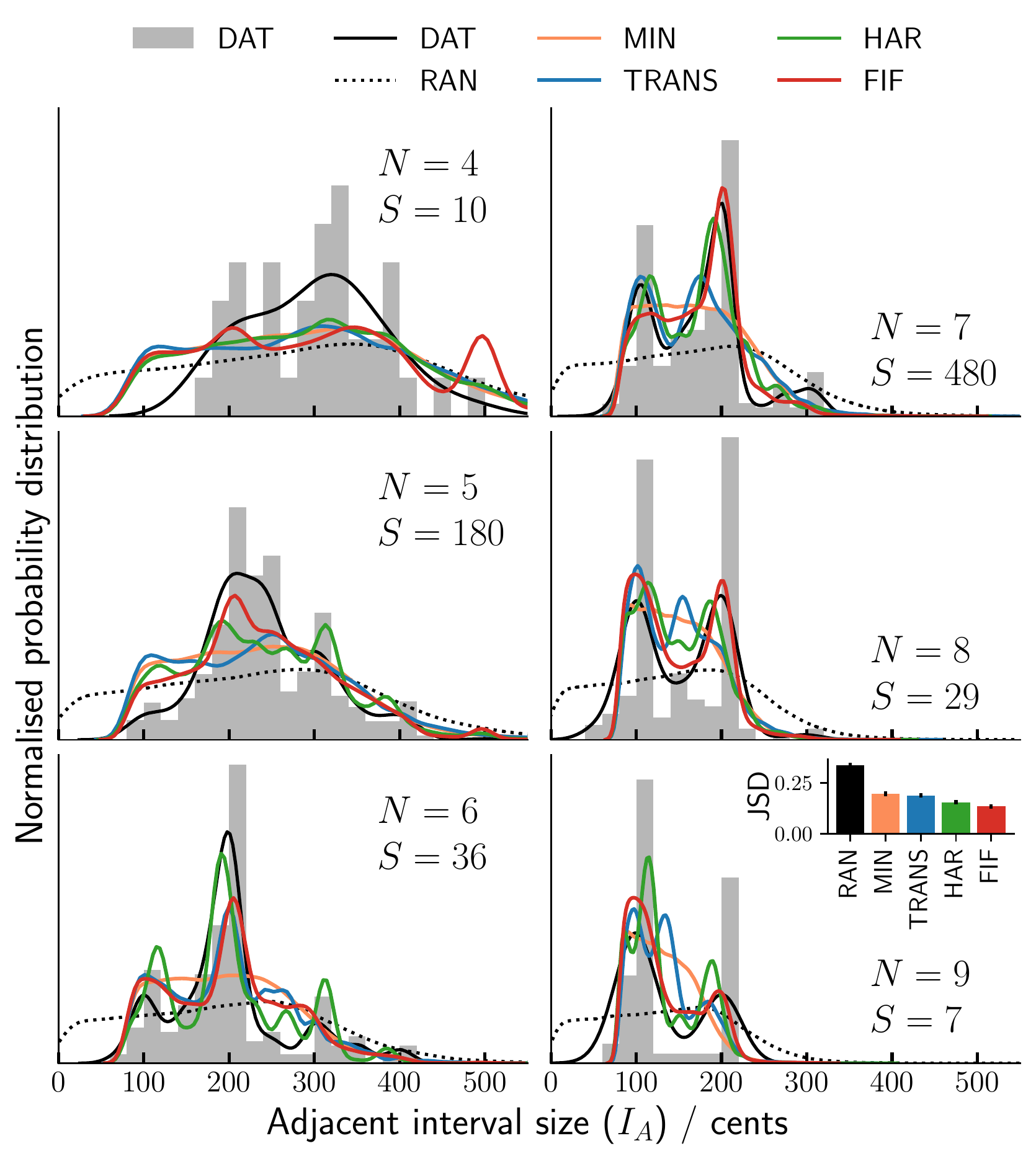}
  \caption{ \label{fig:ia_dist}
  Adjacent interval ($\IP$) probability distributions for empirical \DAT~scales 
  (black solid line) and model scales. The sum of the probabilities is one.
  Lines are fitted to histograms using kernel density estimation.
  Histograms (bin size is $20$ cents) and sample sizes ($S$)
  are shown only for \DAT~scales.
  Each panel shows the distributions for a given note number $N=4-9$.
  Inset: the average Jensen-Shannon divergence, $\textrm{JSD}$, between
  model and \DAT~distributions for each model; whiskers indicate
  $95\%$ confidence intervals.
  }
  \end{figure}

  Adjacent intervals, $\IP$, can be considered the building blocks of scales. 
  Many musical traditions use a restricted set of adjacent intervals
  with which multiple scales are formed.
  Fig. \ref{fig:ia_dist} shows $\IP$ histograms of
  database scales (\DAT) for $N = 4-9$ notes with the sample size, $S$, indicated.
  The $N=5$ and 7 note scales are most common \cite{pat10}.
  However, we have few examples of 4 and 9 note scales, 
  but nonetheless we have included them 
  as they offer some information about the general shape
  of the distributions if not the details.

  The empirical distributions of the \DAT~scales, 
  shown in Fig. \ref{fig:ia_dist}, are clearly not random.
  Some intervals are more common, in particular a peak at 200 cents is
  prominent for all $N$. As $N$ increases, the main peaks in the distributions
  shift to the left (smaller interval sizes) and the range shrinks. 
  This trend holds also for \RAN~scales, 
  and by including the minimum interval constraint (\MIN~model), 
  the ranges of the theoretically generated distributions approximate 
  the empirical ones.
  This indicates that at the most basic level, the choice of intervals
  may be considered a packing problem. Given $N$ intervals of size $\IP$, 
  their distribution depends on the possible ways that these $N$ intervals can be combined into an octave.
  Still, this description fails to explain the significant peaks
  and troughs in the distributions.

\subsubsection*{The \FIF~model best replicates empirical adjacent interval distributions}
  The models which use biases result in distributions with clear peaks and troughs (Fig. \ref{fig:ia_dist}).
  In the \TRANS~distributions, some of the peaks
  are aligned with the \DAT~distributions ($N=6,7$), while others are clearly
  not ($N=8,9$). The existence of prominent peaks at $1200/N$, in addition to other peaks,
  indicates that the bias favours equidistant scales, but not exclusively.
  The \HAR~and \FIF~distributions match many of the features of the
  \DAT~distributions. The sample-size averaged Jensen-Shannon divergence between
  the $\DAT$~and the model distributions (Fig. \ref{fig:ia_dist} inset, SI Table 5)
  show that the \FIF~ model best fits the \DAT~distributions.
  This is somewhat unexpected, as the \FIF~model only specifies the inclusion
  of fifths, a large interval, while it predicts the distribution of smaller
  adjacent intervals. To explain the performance of the \FIF~model,
  we can consider how these building blocks are arranged into scales.

\subsubsection*{The \FIF~model best replicates the empirical note distributions}
\begin{figure}[h]  \centering
  \includegraphics[width=1.00\linewidth]{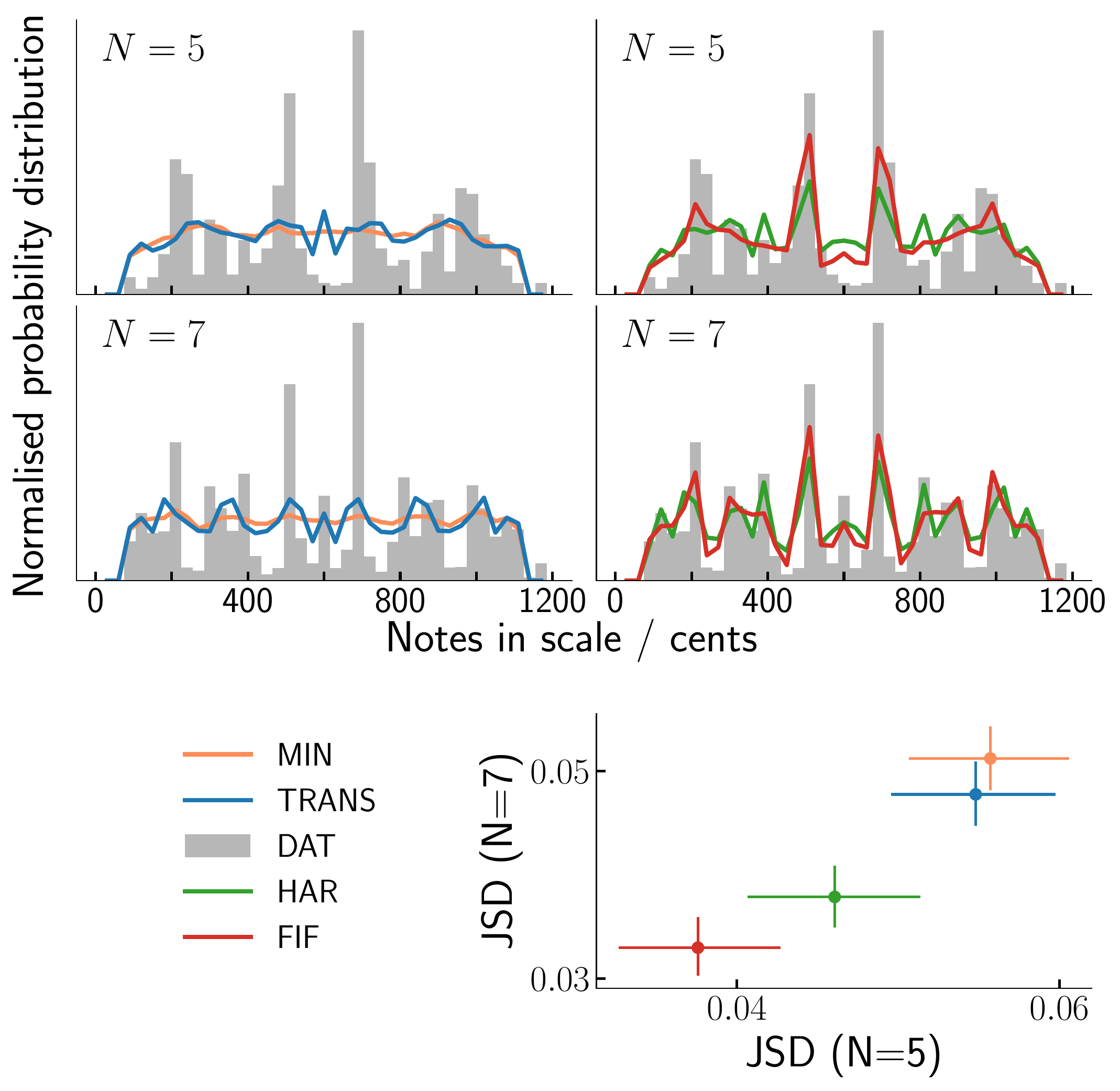}
  \caption{ \label{fig:scale_dist}
  Top: Probability distribution of scale notes for \DAT~and model scales.
  Histograms are shown as lines for model results;
  sample sizes for \DAT~scales are the same as in Fig \ref{fig:ia_dist}.
  Histograms (bin size is $30$ cents) are truncated at $15$ and $1185$ cents for clarity.
  Bottom: JSD between model and \DAT~distributions for $N=5$ and $N=7$;
  whiskers indicate $95\%$ confidence intervals.
  }
  \end{figure}

  The \MIN, and \TRANS~models accept or reject scales based only on their
  adjacent intervals, disregarding their order. Since the \HAR~and \FIF~models
  use all intervals (by comparing all pairs of notes) in their cost functions,
  they take the interval order into account.
  Hence, to see if interval order is indeed an important determinant, 
  we examine how the notes are distributed in the scales (Fig. \ref{fig:scale_dist}).
  Only $N=5$ and $N=7$ are shown due to the stricter need for
  sufficient statistics when considering the full scale distributions. 
  The distributions of \DAT~scales show that certain notes are favoured, 
  with notable peaks at 200, 500 and 700 cents for both $N$. 
  Apart from exclusion zones at the boundaries, every note (when discretized in
  bins of width 30 cents) is used in at least one scale.
  
  Among the models, the \MIN~distributions lack much detail beyond noise, apart from 
  slight undulations with a periodicity of $\frac{1200}{N}$ cents.
  The \TRANS~distributions contain some notable peaks. Peaks at 500 and 700 cents
  fit the \DAT~distributions, while the 600 cents peak for $N=5$ is indubitably
  incorrect. The \HAR~and \FIF~distributions replicate the main features of the \DAT~distributions,
  including peaks, contour and troughs. The extra detail in the \HAR~model
  appears as additional peaks at harmonic intervals in the distributions.

\subsubsection*{Lacking rules for ordering intervals into scales, 
the \TRANS~model still performs well}

\begin{figure}[h]  \centering
  \includegraphics[width=1.00\linewidth]{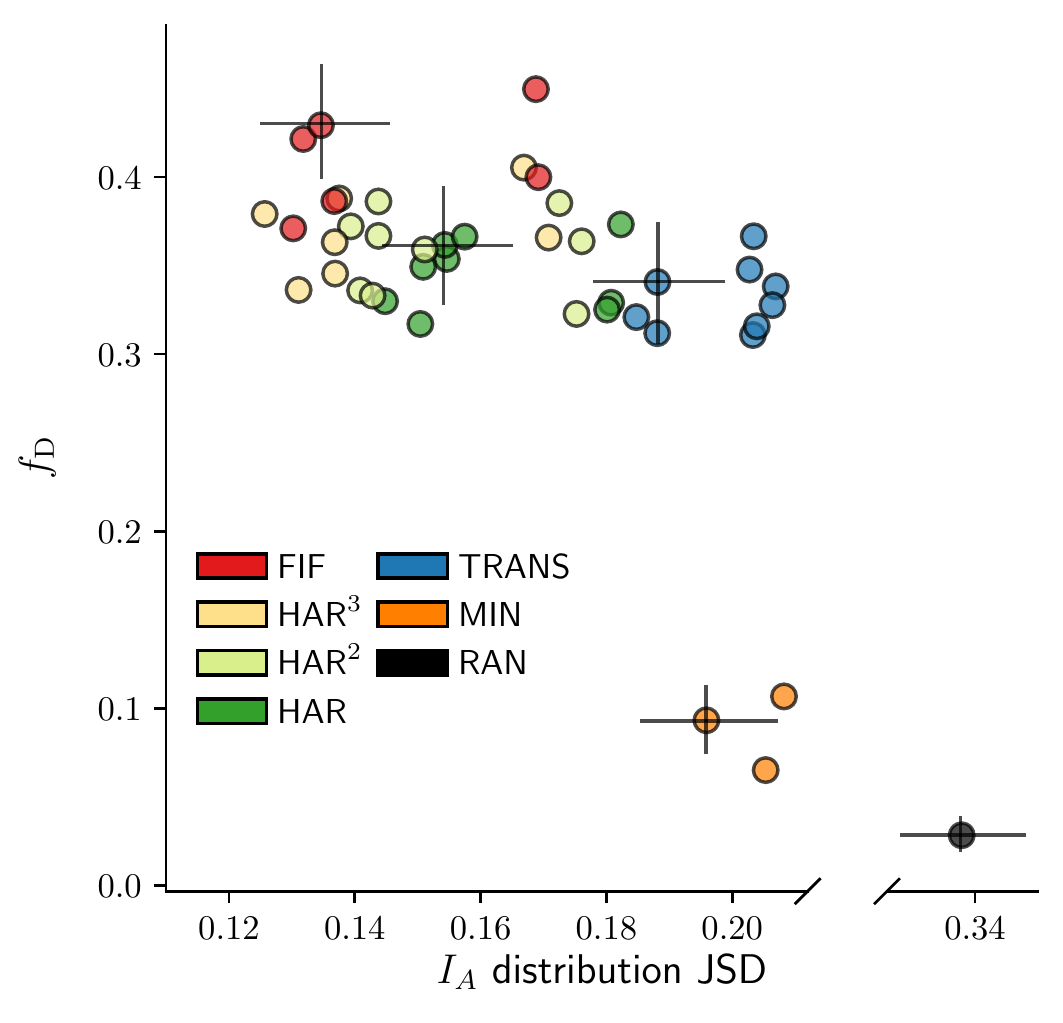}
  \caption{ \label{fig:model_comp}
  Model performance is plotted as the Jensen-Shannon divergence (JSD)
  between the \DAT~and model $\IP$~distributions against $\fD$, the fraction of \DAT~scales predicted.
  Superior models exhibit low JSD values and high values of $\fD$.
  Results for altered versions of the \HAR~model are also shown as \HAR$^2$~and \HAR$^3$.
  For each model, results are shown for combinations of parameters: $\Imin=[70,80,90]$,
  $w=[5,10,15,20]$, $n=[1,2,3]$, and in each case the bias strength $\beta$ is optimized.
  Whiskers indicating $95\%$ confidence intervals are only displayed for the models shown
  in Fig. \ref{fig:ia_dist}, Fig. \ref{fig:scale_dist} and Fig. \ref{fig:cluster}.
  }
  \end{figure}

  We can compactly represent the performance of models by
  two scalar measurables (Fig. \ref{fig:model_comp}):
  \begin{itemize}
      \item[(i)] The sample-size-weighted Jensen-Shannon divergence between \DAT~and theoretical $\IP$ distributions for each model.
      \item[(ii)] $\fD$ -- The fraction of \DAT~scales
  that are found in the model-generated populations.
  \end{itemize}

  Due to our assumption that intervals in a scale add
  up to $1200$ cents, many `measured' scales cannot be found
  by the model due to inexact octave tunings, resulting in an upper
  bound of $\fD\approx0.8$. Additionally, the models generate samples of $S=10^4$
  scales, and while JSD converges by this point, $\fD$ does not -- \ie~as $S$
  increases, so will $\fD$.

  The \TRANS~model is better suited to predicting \DAT~scales
  than predicting \DAT~note distributions. This is surprising, given that the \TRANS~model
  offers no guidance on how to arrange intervals into a scale.
  Correlations between intervals in our database show that two intervals
  are usually ordered such that large goes with small and vice versa (we call this `well-mixed').
  Two small (large) intervals are half as likely to be placed
  together in \DAT~scales than random chance would predict (SI Fig. 3). 
  By shuffling the model scales with a bias towards
  mixing of sizes, we observed an average increase in $\fD$ of $12\%$ for 
  the \TRANS~model while seeing an average decrease of $22\%$ for the
  \FIF~model (SI Fig. 4). So while scales tend to be arranged so that interval sizes
  are well-mixed, it is more important that they are arranged to maximise
  the number of fifths.
  Given that the \FIF~(\HAR) and \TRANS~models are independent of each other,
  the performance of the \TRANS~model is quite notable.

\subsubsection*{\HAR~model performance relies heavily on imperfect fifths}

  \begin{figure*}[h]  \centering
  \includegraphics[width=0.95\linewidth]{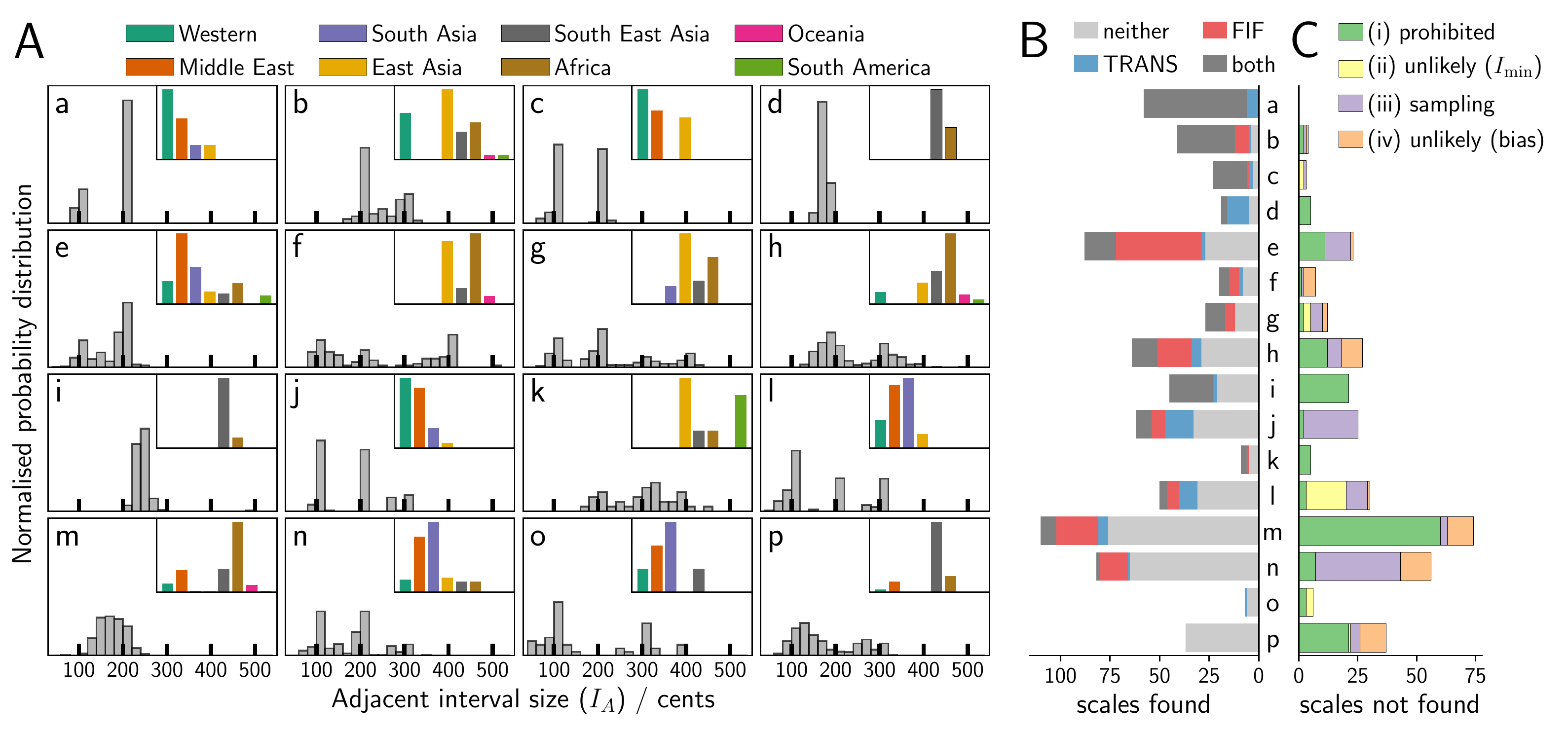}
  \caption{ \label{fig:cluster}
  \textbf{A}: $\IP$ interval distributions of empirical \DAT~scales for
  clusters a-p and their geographical composition (inset).
  \textbf{B}: Number of \DAT~scales found by neither model, only the \TRANS~model,
  only the \FIF~model, or both models, for cluster a-p.
  \textbf{C}: For the scales that are not found by any of the \TRANS, \FIF, or \HAR~models, we show
  the breakdown of the reasons why: (i) prohibited due to model constraints, (ii) unlikely due
  to model constraints, (iii) insufficient sampling, (iv) unlikely due to model biases.
  }
  \end{figure*}

  Despite the \FIF~model being a much simpler version of the \HAR~model,
  the \FIF~model achieves superior results according to our metrics 
  (Fig. \ref{fig:model_comp}). To identify possible reasons, 
  we interpolate between the \HAR~and \FIF~model templates 
  by taking as a harmonicity score the $m$-th power 
  of the harmonicity template in Fig. \ref{fig:model}B, 
  thereby approaching the \FIF~template as $m\to\infty$.
  Fig. \ref{fig:model_comp} shows the results for models where
  we use $m=2$ (\HAR$^2$) and $m=3$ (\HAR$^3$), together with the original 
  \HAR~model ($m=1$).
  The results indicate that as one interpolates from the \HAR~model 
  to the \FIF~model the performance also interpolates continuously.
  Additionally, we note that the \FIF~and \HAR~models are strongly
  correlated (e.g., Pearson's $r=0.76$ at $w=20$, SI Fig. 5B).
  This implies that fifths are a dominant part of the \HAR~model,
  while the extra detail merely hinders performance.

\subsubsection*{Clustering of scales reveals differences between model predictions}

  To further understand how the models work, and in particular why two
  disparate theories (harmonicity and transmittability) perform well, we examined what types of \DAT~scales they
  predict. We divide the scales into 16 clusters by hierarchical clustering
  based on their $\IP$ sets. 
  In Fig. \ref{fig:cluster}A we display the $\IP$ distribution
  and geographical composition for each cluster. Middle Eastern, Western,
  South Asian and East Asian scales appear to have been at one time 
  based on similar mathematics \cite{te70,sig77,far04,kat15,rec18},
  so it might be expected that they should cluster together
  (a, c, j, l).
  There are similarities between some South East Asian and African cultures
  that utilise equidistant 5 and 7-note scales (d, h, i, p).
  Clearly there are also some clusters that might be ill-defined (e, k).

  The number of scales in each cluster found by the 
  \TRANS~and \FIF~models is shown in Fig. \ref{fig:cluster}B.
  For clarity, we omit the \HAR~results as they are qualitatively similar to the
  \FIF~results, but with less scales predicted (SI Fig 6A). The conclusions
  drawn about the differences between the \TRANS~and \FIF~models
  are also relevant when comparing the \TRANS~and \HAR~models.
  All models appear to perform well
  when $\IP$ distributions are uni/bimodal with sharp peaks (a, b, c, i).
  The \TRANS~model performs better if $\IP$ distributions have sharp
  peaks, regardless of how many peaks there are (a, c, d, i, j, l).
  The harmonicity models are better adapted to scales from clusters
  with broad $\IP$ distributions (b, e, f, g, h, m, n).
  Despite these differences, the majority of scales that are found
  are common to all models, and while the \TRANS~and \FIF~(\HAR) cost
  functions are not correlated in \MIN~scales (Pearson's $r=0.01$ ($-0.04$))
  they are correlated in \DAT~scales (Pearson's $r=0.41$ ($0.12$)) (SI Fig 7).
  Ultimately, of those scales that are predicted, a majority appear to be selected
  for both harmonicity and transmittability.

\subsubsection*{A majority of scales were predicted by at least one model}

  Between the three models, $407$ out of $742$ ($55\%$) scales were found.
  There are several reasons why the other scales were not found:
  (i) It is impossible for the model to generate them due to having
  an interval smaller than $\Imin$ or deviations from a perfect octave.
  (ii) It is possible but unlikely for the model to generate them
  due to having intervals at the tails of the $\IP$ distributions
  - this is a feature of the random sampling, which is compounded
  by the use of an $\Imin$ constraint.
  (iii) They are likely to be found, but we generated too few samples
  to find them all.
  (iv) They are not likely to be found by any model.

  We calculated the breakdown of the reasons why scales were not found
  (Fig. \ref{fig:cluster}C). Scales are in category (i)
  if $\Imin<70$ or if the octave deviates from $1200$ cents by more than
  $10$ cents ($155$ scales). For all scales we calculate
  the probability, $P_{\MIN}$, that they are predicted after applying the $\Imin$ constraint,
  and note the value of $P_{\MIN}$ below which only $10\%$ of scales are not found.
  Scales with $P_{\MIN}$ lower than this value are in category (ii) ($26$ scales), or in one
  of categories (iii) and (iv) otherwise.
  We calculate for all scales the probability, $P_{\ANY}$,
  that a scale will be found by any model, and note the value of $P_{\ANY}$
  below which only $10\%$ of scales are not found. Scales with $P_{\ANY}$ higher than
  this value are in category (iii) ($100$ scales), otherwise (iv) ($54$ scales).
  When we only consider scales that the model can predict (categories ii-iv),
  we find that the three models are not particularly sensitive to `theory' or
  `measured' scales, nor the region or culture from which they come (SI Fig 6).
  This implies that these biases are in effect consistently across musical traditions.
   
\section*{Discussion}

\subsubsection*{Packing arguments account for the success of the harmonicity models}

  The \FIF~model manages to reproduce the main features of the interval ($\IP$)
  and scale distributions, simply by specifying inclusion of fifths. 
  This may seem counter-intuitive, but it can be rationalised entirely
  in terms of packing. Consider first Fig. \ref{fig:scale_dist}. 
  Since $0$ and $1200$ cents are fixed points, the easiest way to add
  fifths is by having notes at $500$ ($1700$) or $700$ cents. 
  Taking into account the circular nature of scales, 
  one can subsequently add notes at $200$ ($1400$) cents or $1000$ cents,
  which will further interact with previously added notes to create more fifths.
  As a result, there are exclusion zones at either side of $500$ and
  $700$ cents. Packing also explains the features of the $\IP$ distributions
  in Fig. \ref{fig:ia_dist}: if $200$, $500$, $700$ 
  and $1000$ cents are all highly probable, then the $\IP$ distributions ought to contain
  many $200$ cents intervals regardless of $N$.

  It is likely that packing is also the main mechanism in the \HAR~model.
  The \HAR~model is affected not only by the harmonic template
  (Fig. \ref{fig:model}B), but also the constraints of fixed $N$ and octave size.
  The \HAR~results then arise from the ways of
  arranging a scale to maximize its average harmonicity score.
  The fifth is the most important interval in the \HAR~model, as evidenced
  by its high score and also the high correlation between
  the \FIF~and \HAR~cost functions (SI Fig 5B).
  The similarities between the \FIF~and \HAR~results
  (Fig. \ref{fig:ia_dist}, \ref{fig:scale_dist} and SI Fig. 6A)
  provides additional evidence that the \HAR~model shares the \FIF~packing mechanism.
  Constructing scales from fifths is an old concept \cite{wesml94}.
  The new understanding is that different cultures can be explained
  in this way even when there is no evidence that they explicitly
  tuned instruments using fifths (SI Fig. 6).

\subsubsection*{The devil in music? -- The devil is in the detail}
  The tritone interval (${\sim}600$ cents) has been
  traditionally considered dissonant in Western music -- 
  earning it the name \textit{diabolus in musica} \cite{tre18} -- 
  and is uncommon in classical and folk music \cite{vosmp89}.
  Is it rare because it is unpleasant, or unpleasant because it is rare?
  Viewing scales as a packing problem reveals an alternative
  explanation:
  as $N$ increases, the average $\IP$ size decreases,
  and it becomes easier to simultaneously pack both tritones and fifths.
  Analysis of the database shows a linear relationship between $N$
  and the frequency of tritone intervals, and this trend is replicated
  best by the \FIF~model (SI Fig. 8).
  Therefore according to this theory, the tritone is rarely used in
  music simply because it is difficult to simultaneously pack tritones and fifths.

\subsubsection*{On the universality of harmonic intervals}
  The oft-repeated claim that harmonic intervals are universal is based on
  sparse statistical evidence \cite{tre18}. Our data shows that harmonicity and
  prevalence are correlated, and this correlation depends on $N$
  ($N=5$, $r=0.70$, $p<0.005$; $N=6$, $r=0.43$, $p>0.05$;
  $N=7$, $r=0.65$, $p<0.005$; SI Fig. 9). Fifths ($3/2$, $702$ cents) and,
  due to inversion, fourths ($3/2$, $498$ cents) are the
  most widespread. However the ratios $7/6$ ($267$ cents)
  and $7/4$ ($968$ cents) are quite rare despite their relatively high harmonicity,
  in contrast to $11/7$ ($783$ cents) which is more common yet less harmonic.
  Thus we report the first clear
  evidence for prevalence of harmonic intervals, showing that the
  correlation between prevalence and harmonicity is not straightforward.

\subsubsection*{Multiple selection pressures affect evolution of scales}
  All three models, \HAR, \FIF~and \TRANS, explain the empirical data significantly
  better than chance. Thus it is possible that scales are simultaneously optimized
  for compressibility, maximization of fifths, and maximization of other harmonic
  intervals. Out of the two harmonicity models, the \FIF~model better fits the data than the \HAR~model,
  but we note that these models are the two extreme cases in how they treat
  higher order harmonics. A more sophisticated model may have a nuanced
  approach that accounts for the spectra of natural sounds and how
  they are processed. There is also a speculative scenario in which
  harmonic intervals are selected through a different mechanism.
  The cultures which use small-integer frequency ratios (`theory'
  scales) overlap with those that had contributed to the development of mathematics \cite{go16}.
  Thus the use of mathematics may have contributed to the development
  of scales \cite{croja63}.

  A significant minority of scales are not supported
  by any model. A view some may consider natural is that
  perhaps no one mathematical model can predict the diversity of scales.
  Regarding the theories tested:
  cultures which produce monophonic music may ignore harmony \cite{mcdna16};
  cultures which primarily focus on rhythm and dance may
  not mind imprecision in melodies. 
  Consider the Gamelan pelog scale, 
  variations of which account for over two thirds of cluster p. 
  Pelog scales rarely contain imperfect fifths (SI Fig. 10A).
  A crude approximation of the pelog scale
  reduces it to intervals on a 9-TET (9-tone equal temperament) scale \cite{rahyi78}.
  This scale is composed of 5 small intervals (average size 133 cents)
  and 2 large intervals (average size 267 cents). 
  This should be predicted by the \TRANS~model, 
  but within individual pelog scales the deviations from
  9-TET are so large that this rarely happens.
  
  The pelog scale is not exceptional in this regard;
  huge deviations from the average are seen 
  in Thai scales and the Gamelan slendro scale (SI Fig. 10).
  In this case, some propose that the deviations from some
  mathematical ideal are not examples of mistuning, 
  but rather a display of artistic intent \cite{vetet89,maram93,garan15}.
  Tuning as a means of expression is rare in the West, 
  but not unheard of in classical or popular music 
  (see La Monte Young's "The Well-tuned Piano" 
  or King Gizzard and the Wizard Lizard's "Flying Microtonal Banana").
  While these theories have performed well, 
  the variability in scales of some cultures 
  hints that although scales are inherently mathematical, 
  mathematics alone lacks the power to fully explain our choices of scales.

\subsubsection*{Limitations due to assumptions and data}
  We report that the \FIF~model best fits the available data,
  and we assume that the data is representative
  of scales used in different musical traditions. However, a different data collection
  method may significantly alter the results. We replicated the analysis
  with sub-samples of our database and found that the conclusions
  are at least robust to resampling (SI Fig 11). Future work may benefit
  from controlling for historical relationships between musical traditions \cite{savpn15}.

  Despite evidence that tonal hierarchies are common, they are not
  a necessary element of scales so we avoid assumptions
  about tonality \cite{mcdmp05, kesmp84}. We could have assumed
  that the tonic (the note at $0$ cents) is special, and so
  are intervals made with the tonic. This would explain why in Fig. 4
  the $700$ cents is more salient than $500$ cents.
  We note that the we could have chosen a different proxy measure of
  harmonicity \cite{harpr19}. However, we think that doing
  so would not alter the conclusions as the different measures 
  are highly correlated (SI Fig 5A).

\subsubsection*{Lack of hexatonic scales may be owing to historical convention}
  
  Why do we choose $N$ notes in a scale? --
  It is generally agreed that there is some trade-off involved.
  With too few notes melodies lack complexity; with too many
  notes melodies are difficult to learn.
  Too few notes results in larger intervals which are more difficult
  to sing \cite{sunjv93}; too many notes results 
  in smaller intervals which have lower harmonic similarity \cite{gilpo09}.
  This simple trade-off is at odds with the contrasting ubiquity of
  5 and 7 note scales and scarcity of 6 note scales.
  One suggestion is that 6 note scales are actually prevalent, but classified
  as variants of 5 and 7 note scales due to convention.
  Evidence for this can be found in the Essen folk song collection \cite{sch95}.
  
  Chinese music and Western music are conventionally thought to be 
  composed using 5 and 7 note scales respectively, but for both 
  cultures, 6 note scales are the second most prevalent in folk songs (SI Fig. 12).
  While the following is mere speculation, this work points to
  a possible route by which a preference for pentatonic and heptatonic
  scales may have arisen. Given the simplicity of equidistant scales,
  they are the easiest compressible scales to evolve.
  For 5, 6, 7 and 8 note equidistant scales, 
  the closest notes to fifths are 720 cents, 600 (800) cents, 686 cents, and
  750 cents, respectively.
  Thus, 5 and 7 note scales are the only equidistant scales
  that can include imperfect fifths.
  The paucity of recorded hexatonic scales may therefore be due to historical
  convention driven by evolutionary pressures in early music.

\section*{Conclusion}

  By constructing a cross-cultural database and using 
  generative stochastic modelling, 
  we quantitatively tested several theories on the origin of scales. 
  Scales tend to include imperfect fifths, 
  and the features of the empirical distributions arise 
  from the most probable ways of packing fifths into a scale of fixed length.
  Scales also tend to be compressible, which we suggest
  leads to melodies that are easier to communicate and remember.
  There is evidence that efficient data compression is 
  a general mechanism humans use for discretizing continuous signals.
  The effect of compressibility on pitch processing merits further research.
 
  No single theory could explain 
  the diversity exhibited in the database.
  It is instead likely that scales evolved subject 
  to different selection pressures across cultures. 
  These theories can be further tested by expanding the database,
  in particular by computationally identifying scales used in ethnographic
  recordings. 
  Out of two harmonicity-based theories, the one that best fits the empirical
  data only considers the first three harmonics.
  This may shed some light on the important, but still developing, 
  understanding of how harmonicity is processed in the brain.

\section*{Methods}
\subsubsection*{The stochastic models}
  The stochastic models generate adjacent intervals $\IP$ from a uniform distribution,
  which are then scaled so that they sum to $1200$ cents. 
  Some models apply a minimum interval constraint, $\Imin$, 
  such that no $\IP$ is smaller than $\Imin$.
  Depending on the model, scales are accepted or rejected according to a probability
  \begin{equation}
  P = \min \left\{ 1, \exp( - \beta C) \right\}~,
  \end{equation}
  \noindent where $C$ is a cost function, and we tune the strength
  of the bias with the parameter $\beta$. We tested
  many cost functions to check that the results are insensitive to the
  exact functional form (SI Fig. 13, 14, 15).
  
  For the harmonicity model \HAR,
  \begin{equation}
  C_{\HAR} = 1 - \frac{\bar H - \bar H_{min}}{\bar H_{max} - \bar H_{min}}~,
  \end{equation}
  \noindent where $\bar H_{min}$ and $\bar H_{max}$ are normalization constants
  (these constants, obtained via random sampling, are listed in SI Table 4),
  and the average harmonicity score $\bar H$ is given by
  \begin{equation}
  \bar H = \bigg[ N(N-1) \bigg]^{-1} \sum^{N-1}_{i=0} \sum^{i+N-1}_{j=i+1} H(I_{ij}, w)~,
  \end{equation}
  \noindent where $H(I)$ is the harmonic similarity score 
  of an interval size $I$, $w$ is the size of the window 
  in which intervals are considered equivalent,
  and $I_{ij}$ is the interval between note $i$ and note $j$. 
  The index $i=0$ refers to the starting note of the scale 
  and $j$ takes into account the circular nature of scales
  (if $j>N$ then note $j$ is an octave higher than note $j-N$).
  Note that we do not consider either unison or octave intervals in this score.
  The harmonic similarity score $H$ of an interval is calculated 
  from its frequency ratio expressed as a fraction.
  \begin{equation}
  H(I) = \frac{x+y-1}{xy}\times100~,
  \end{equation}
  \noindent where $x$ is the numerator and $y$ is the denominator of the fraction.
  The harmonic similarity template is produced by creating a grid 
  of windows of maximum size $w$ centred about the intervals 
  that have the largest $H$ values. An interval expressed in cents
  is allocated to the window with the highest $H$ value that is within $w/2$ cents.
  
  For the imperfect fifths model \FIF,
  \begin{equation}
  C_{\FIF} = ( 1 + \bar F )^{-1}~,
  \end{equation}
  \noindent where a constant is added to prevent division by zero 
  in the case of scales with no imperfect fifths, and 
  the fraction of fifths $\bar F$ is
  \begin{equation}
  \bar F = \bigg[ N(N-1) \bigg]^{-1} \sum^{N-1}_{i=0} \sum^{i+N-1}_{j=i+1} F(I_{ij}, w)~,
  \end{equation}
  \noindent where $F(I,w)=1$ if $|I-702| \leq w/2$ and $0$ otherwise.
  
  For the transmittable scales model \TRANS,
  \begin{equation}
  C_{\TRANS} = \min_{\frac{1}{2}\min\{\IP\} \leq \gamma \leq \frac{3}{2}\min\{\IP\}} 
  \left\{ \sum^{N}_{i=1}
  \Bigg| \left\lceil \frac{I_{Ai}}{\gamma} \right\rceil - 
  \frac{I_{Ai}}{\gamma} \Bigg| ^n \right\} / N~,
  \end{equation}
  \noindent where $I_{Ai}$ is the $i^{\rm th}$ adjacent interval in a scale, 
  and $n$ is parameter that controls how deviations from the template 
  are considered in the model (see main text). 
  The parameter $\gamma$ is   the common denominator 
  of the compressible template; it is constrained so that it
  is never less than half of the smallest adjacent interval.

\subsubsection*{Classification of scales as similar}
  The fraction of \DAT~scales $\fD$ for each model was calculated
  by checking if model scales are similar to \DAT~scales.
  Two scales A and B are similar if
  \begin{equation}
  \forall i \in \{1..N\}, | \alpha^A_i - \alpha^B_i | \leq e ~,
  \end{equation}
  \noindent  where $\alpha^A_i$ is $i^{th}$ note in scale $A$ 
  and the tolerance is $e=10$.
  
\subsubsection*{Probability of predicting scales}
  $P_{\MIN}$, the probability of a scale A being predicted by the \MIN~model,
  is calculated as the sum of the probabilities of each scale $B$ 
  that can be labelled similar to, or the same as, scale $A$. 
  We use a tolerance of $e=10$, we keep the length of the scale
  constant, and we consider probabilities at an integer resolution, 
  so there are $21^{N-1}$ similar scales.
  \begin{equation}
  P_{\MIN} = \sum^{21^{N-1}} \prod^{N}_{i=1} p(I^B_{\textrm{A}i})~,
  \end{equation}
  \noindent where $p$ is the probability of an interval being picked
  by the \MIN~model, and $I^B_{\textrm{A}i}$ is the $i^{th}$ adjacent interval in scale B.

  $P_{\ANY}$, the probability that any of the \TRANS, \HAR~or \FIF~models
  finds a scale is given by
  \begin{equation}
  P_{\ANY} = P_{\MIN} \sum P(\beta, C)~,
  \end{equation}
  \noindent where $P$ is summed over the three models using the parameters
  listed in SI Table 4.

\subsubsection*{Clustering criterion}
  To group \DAT~scales we used hierarchical clustering 
  based on distances between adjacent intervals. 
  The distance, $d_{AB}$, from scale $A$ to scale $B$ is asymmetric, 
  and is calculated as the sum of the shortest distances 
  from every $\IP$ in $A$ to any $\IP$ in B,
  \begin{equation}
  d_{AB}  = \sum_{i=1}^{N} \min_{j} \{I^A_{\textrm{A}i} - I^B_{\textrm{A}j}\}~,
  \end{equation}
  where $i$ and $j$ are the indices of adjacent intervals 
  in scales $A$ and $B$ respectively. 
  For clustering, we use the symmetric distance
  $d_C = (d_{AB}d_{BA})^{1/2}$. 
  We used Ward's minimum variance method to agglomerate clusters 
  using the SciPy package for Python \cite{jon01}.

\subsubsection*{Statistical analysis}

  We tested whether the empirical $\IP$ and note distributions
  are better approximated by a theoretical model (\FIF,\HAR,\TRANS),
  compared to the \RAN, and the \MIN~model, which
  are effectively the null distributions.
  We chose to smooth the $\IP$ distributions in Fig. \ref{fig:ia_dist}
  mainly due to one assumption. Given that (with the exception of
  modern fixed pitch instruments) intervals are not produced as exact
  frequency ratios, the $\IP$ distributions should be smooth.
  We consider that the sharp peaks (\eg at $100$ and $200$ cents are
  a result of theoretical values for interval sizes and are not representative
  of how intervals are produced in reality. To minimize artefacts
  from this smoothing we used a non-parametric kernel density estimation
  method, implemented in the Statsmodels package for Python \cite{sea10}.
  The kernel is Gaussian and Silverman's rule is used to estimate the bandwidth.
  We compared goodness-of-fit between the empirical and theoretical
  distributions using the Jensen-Shannon divergence (we get the arithmetic mean
  across $N$, weighted by sample size). We verified that the smoothing did not influence the results
  by using a two sample Cram\'er-von Mises test (SI Table 5). 
  Due to the high dimensionality of the system we needed to generate large
  samples ($S=10^4$) for each model. As a result, $p$ values tend to be
  astronomically low ($p \ll 0.01$), so we instead calculated $95\%$
  confidence intervals by bootstrapping (1000 resamples from the \DAT~sample).
  Applying kernel density estimation to the note distributions
  (Fig. \ref{fig:scale_dist}) resulted in a loss of detail via oversmoothing.
  Hence, we only show histograms for these distributions and calculate the
  Jensen-Shannon divergence using the histograms. The conclusions do
  not depend on the histogram bin size.
  To test whether we have sufficient empirical data we replicate the analysis with three
  types of sub-sampling: bootstrapping with smaller sample sizes; only the `theory' scales;
  and only the `measured' scales (SI Fig 11). We find that the
  conclusions are robust to resampling.

\subsection*{Data and Code Availability} 
  All data used in the figures and Supplementary Information are available,
  along with simulation and analysis code are accessible at
  \url{https://github.com/jomimc/imperfect_fifths}.
  The scales database is included as Supplementary Material.

\subsection*{Acknowledgements} 
  J.M. acknowledges helpful discussions with Patrick Savage.
  This work was supported by the taxpayers of South Korea through the
  Institute for Basic Science, Project Code IBS-R020-D1.

\subsection*{Competing Interests} The authors declare that they have no
  competing financial interests.

\subsection*{Correspondence} Correspondence and requests for materials
  should be addressed to J.M.~(email: jmmcbride@protonmail.com)
  and T.T.~(tsvitlusty@gmail.com).

\subsection*{Author Contributions}
  J.M. and T.T. designed research; J.M. performed research;
  J.M. analyzed data; J.M. and T.T. wrote the paper.

\bibliographystyle{unsrtnat}
\bibliography{scales_paper}


\end{document}


\maketitle

\subsection*{Prominent intervals in the harmonic series}

  We estimate the prominence of intervals that are made between harmonics
  in a single harmonic series by weighted counting. We count the intervals made 
  between the first $n_1$ harmonics and the first $n_2$ harmonics, with each
  interval combination only counted once. We consider
  two cases: $n_1=1$ and $n_2=10$; $n_1=10$ and $n_2=10$. In the first case,
  only the intervals made with respect to the fundamental are counted. In the
  second case, all possible intervals within a harmonic series are counted, up
  to a maximum harmonic number $10$. We weight the intervals with $\lambda = (1-a)^{i-1}(1-a)^{j-1}$,
  where $a$ is the attenuation rate, 
  such that the weights decay with harmonic number following a power law.
  We find for both cases that as the attenuation rate increases the 
  relative counts of intervals diverge (Table \ref{tab:har_ints}). E.g., for
  the second case with $a=0.3$, the ratios between the
  weighted counts for unison, octave, fifth and major third are approximately $4:4:2:1$;
  while with $a=0.6$ they are $25:5:2:1$.

\begin{table}\centering
\caption{\label{tab:har_ints}
  The most prominent harmonics present in a single harmonic series
  of length $n_2=10$, as calculated by weighted counting.
  $n_1=1$ corresponds to only counting intervals made with respect
  to the tonic, while $n_1=10$ corresponds to counting all possible intervals.
  $a$ is the attenuation rate: $a=0$ means that all harmonics have
  equal weights; $a=1$ means that only the first harmonic is counted.
  }
\begin{tabular}{l|cc|cc|cc|cc|cc|cc}
 $n_1$ &  \multicolumn{6}{c|}{1}  &  \multicolumn{6}{c}{10}  \\ 
 $a$  &  \multicolumn{2}{c|}{0}  &  \multicolumn{2}{c|}{0.3}  &  \multicolumn{2}{c|}{0.6}  &  \multicolumn{2}{c|}{0}  &  \multicolumn{2}{c|}{0.3}  &  \multicolumn{2}{c}{0.6}  \\ 
 & ratio & count & ratio & count & ratio & count & ratio & count & ratio & count & ratio & count  \\ 
\toprule
  &  2/1 & 3.0 & 2/1 & 1.0 & 1/1 & 1.0 & 1/1 & 10.0 & 1/1 & 2.0 & 1/1 & 1.0  \\ 
  &  5/4 & 2.0 & 1/1 & 1.0 & 2/1 & 0.5 & 2/1 & 8.0 & 2/1 & 2.0 & 2/1 & 0.5  \\ 
  &  3/2 & 2.0 & 3/2 & 0.7 & 3/2 & 0.2 & 5/4 & 7.0 & 3/2 & 1.0 & 3/2 & 0.2  \\ 
  &  7/4 & 1.0 & 5/4 & 0.3 & 5/4 & 0.03 & 3/2 & 7.0 & 5/4 & 0.6 & 5/4 & 0.04  \\ 
  &  1/1 & 1.0 & 7/4 & 0.1 & 7/4 & 0.004 & 9/8 & 4.0 & 7/4 & 0.2 & 4/3 & 0.01  \\ 
  &  9/8 & 1.0 & 9/8 & 0.06 & 9/8 & 0.0007 & 7/4 & 3.0 & 4/3 & 0.2 & 7/4 & 0.006  \\ 
  &   &  &  &  &  &  & 4/3 & 3.0 & 5/3 & 0.1 & 5/3 & 0.004  \\ 
  &   &  &  &  &  &  & 5/3 & 3.0 & 9/8 & 0.1 & 9/8 & 0.001  \\ 
  &   &  &  &  &  &  & 7/6 & 2.0 & 7/6 & 0.08 & 7/6 & 0.0007  \\ 
  &   &  &  &  &  &  & 8/5 & 1.0 & 6/5 & 0.04 & 6/5 & 0.0003  \\ 
\end{tabular}
\end{table}

\subsection*{Scales database and inclusion criteria}

  Scales are obtained from three types of sources. Scales with fixed theoretical
  interval sizes (labelled `theory' scales in the main text)
  are obtained from books \cite{hew13, rec18}. A separate set of scales
  (labelled `measured' scales) come from two types of sources:
  scales which are inferred from measurements of instrument tunings
  \cite{wacna50,kubam64,traam70,traam71,sur72,mor76,haee79,kubam80,
   vanam80,zeme81,anija82,lutam82,kubam84,kubyt85,got85,yuapp87,keemp91,traam91,
   carja93,sche01,attpi04,zhaa04,set05,lipi06,mcnja08,kus10,wigjp11,
   garan15,wis15,moric18,bad19};
  scales which are inferred from analysis of recorded musical performances
  \cite{kim76,falae13}. A full breakdown is provided in Table \ref{tab:data_source}.
  
  All of the `theory' scales are included at least once in the database, and
  in many cases more than once if is known that it was used in multiple tuning
  systems (Table \ref{tab:tunings}). In addition, scales were included
  in the database according to the following criteria:
  \begin{itemize}[itemsep=-2pt]
  \item Scales must be reported in sources as intervals in cents or as frequency ratios of notes.
  \item Only scales with at least 4 and at most 9 notes are included.
  \item If a scale has notes beyond the octave these notes are excluded.
  \item Scales within a single culture that are identical within 1 cents tolerance are only included once
            (\textit{i.e.} key changes are not considered).
  \item When the source omits information about the tonic, the tonic is taken as the first
        note for which a consecutive series of intervals add up to within 50 cents of an octave.
        Those intervals are only used in one scale (\textit{i.e}. `modes' are not inferred from `measured' scales)
  \item A number of sources lack the exact value of the final interval
        \cite{got85, wigjp11, falae13}. If there is evidence that the scale includes
        the octave, then a final interval is appended so that the scale ends on the octave.
  \item Scales are excluded if there are significant inconsistencies or errors in the reporting.
  \end{itemize}

\begin{table}\centering
\caption{\label{tab:data_source} The number of scales of each type from each source.
Theory scales have exact frequency ratios specified for each interval.
Instrument scales are inferred from measurements of instrument tunings.
Recording scales are inferred computationally from recordings. }
\begin{tabular}{l|cccc}
 Reference & Theory & Instrument & Recording & Total   \\ 
\toprule
\cite{rec18} & 176 & 67 & 0 & 243  \\ 
\cite{hew13} & 228 & 0 & 0 & 228  \\ 
\cite{sur72} & 0 & 51 & 0 & 51  \\ 
\cite{vanam80} & 0 & 46 & 0 & 46  \\ 
\cite{falae13} & 0 & 0 & 28 & 28  \\ 
\cite{lutam82} & 21 & 0 & 0 & 21  \\ 
\cite{garan15} (raw data obtained via personal communication) & 0 & 17 & 0 & 17  \\ 
\cite{set05} & 0 & 12 & 0 & 12  \\ 
\cite{mcnja08} & 0 & 9 & 0 & 9  \\ 
\cite{zeme81} & 0 & 7 & 0 & 7  \\ 
\cite{yuapp87} & 0 & 7 & 0 & 7  \\ 
\cite{haee79} & 0 & 6 & 0 & 6  \\ 
\cite{kubam64} & 0 & 6 & 0 & 6  \\ 
\cite{kubam80} & 0 & 5 & 0 & 5  \\ 
\cite{bad19} & 0 & 5 & 0 & 5  \\ 
\cite{keemp91} & 0 & 5 & 0 & 5  \\ 
\cite{zhaa04} & 0 & 4 & 0 & 4  \\ 
\cite{kim76} & 0 & 0 & 4 & 4  \\ 
\cite{kus10} & 0 & 4 & 0 & 4  \\ 
\cite{wis15} & 0 & 4 & 0 & 4  \\ 
\cite{sche01} & 0 & 4 & 0 & 4  \\ 
\cite{anija82} & 0 & 4 & 0 & 4  \\ 
\cite{kubam84} & 0 & 4 & 0 & 4  \\ 
\cite{got85} & 0 & 3 & 0 & 3  \\ 
\cite{traam70} & 0 & 3 & 0 & 3  \\ 
\cite{carja93} & 0 & 2 & 0 & 2  \\ 
\cite{wigjp11} & 0 & 2 & 0 & 2  \\ 
\cite{kubyt85} & 0 & 2 & 0 & 2  \\ 
\cite{mor76} & 0 & 2 & 0 & 2  \\ 
\cite{traam91} & 0 & 2 & 0 & 2  \\ 
\cite{wacna50} & 0 & 2 & 0 & 2  \\ 
\cite{traam71} & 0 & 2 & 0 & 2  \\ 
\cite{moric18} & 0 & 1 & 0 & 1  \\ 
\cite{attpi04} & 0 & 1 & 0 & 1  \\ 
\end{tabular}
\end{table}

\begin{table}\centering
\caption{\label{tab:tunings}
  The tunings used for `theory' scales depending on culture.
}
\begin{tabular}{lcccccccc}
  Culture &  12-tet & 24-tet & 53-tet & Just Intonation & Pythagorean & Persian & Turkish & Shi-er-lu  \\
\midrule
  Western Classical (years 1700+) & X &   &   &   &   &   &   &   \\
  Jazz                      & X &   &   &   &   &   &   &   \\
  Diatonic modes            & X &   &   & X & X &   &   &   \\
  Greek Folk                & X &   &   & X &   &   &   &   \\
  Jewish                    & X &   &   & X &   &   &   &   \\
  Japanese                  & X &   &   &   & X &   &   &   \\
  Chinese                   & X &   &   &   &   &   &   & X \\
  Hindustani                & X &   &   & X &   &   &   &   \\
  Carnatic                  &   &   &   & X &   &   &   &   \\
  Arabian                   &   & X & X &   &   &   &   &   \\
  Persian                   &   &   &   &   &   & X &   &   \\
  Turkish                   &   &   & X &   &   &   & X &   \\
\bottomrule
\end{tabular}
\end{table}

\subsection*{Vocal mistuning theory predicts a minimum interval size}
  The vocal mistuning theory states that both singing and pitch perception
  are subject to errors, and hence scales evolved so that they were
  easy to sing and hear \cite{pfojc17}. It is often assumed that error distributions 
  are likely to be Gaussian \cite{siepp77,burja94,seris11,larbr19}.
  By modelling this phenomenon as a transmission
  problem we assess the effect of interval size on errors in transmission.
  That is, given Gaussian errors on both interval production and
  perception, what is the probability that a sung interval will be
  perceived as it is intended? 

\begin{figure}
\centering
\includegraphics[width=\textwidth]{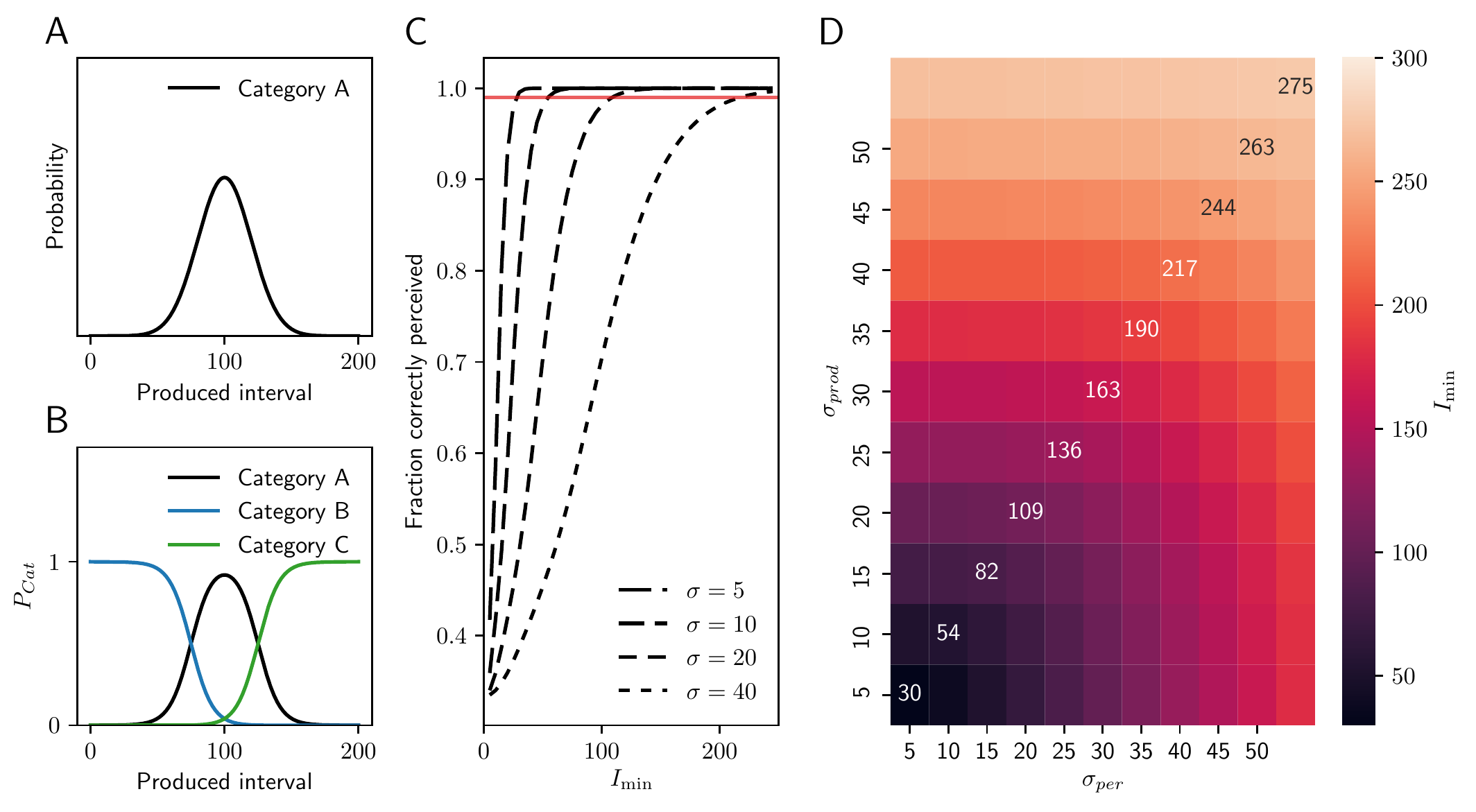}
\caption{\label{fig:vmt_model}
  A: Probability distribution of a sung interval which is intended to be $100$ cents;
     $\mu = 100$ cents and $\sigma=20$ cents.
  B: Probability that a sung interval will be heard as a category A, B or C;
     $\mu = [50, 100, 150]$ and $\sigma=20$.
  C: The fraction of intervals that are perceived as intended as a function
     of the smallest separation between categories, $\Imin$;
     the red line indicates the point at which $99\%$ of intervals are correctly
     perceived; $\sigma = \sigmaprod = \sigmaper$.
  D: The minimum interval size $\Imin$ for which $99\%$ of intervals are perceived
     correctly, as a function of $\sigmaper$ and $\sigmaprod$.
}
\end{figure}

  We vary the standard deviation of
  the size of the produced interval, $\sigmaprod$, and the standard
  deviation of the probability that an interval $I$ will be perceived as
  a specific category, $\sigmaper$. The means of
  interval categories are separated by a distance $\Imin$.
  The probability, $P_{\rm Cat}$, that an interval $I$ will be perceived as a category
  A is
  \begin{equation}
  P_{\rm Cat} = 
      \exp \left( -\frac{(I-\mu_{\textrm{A}})^2}{2 \sigmaper^2} \right)
      \left[\sum_X {
      \exp \left( -\frac{(I-\mu_{\textrm{X}})^2}{2 \sigmaper^2} \right)} \right]^{-1}~,
  \end{equation}
  \noindent where $\mu_{\textrm{X}}$ is the mean of any category X.
  Thus we can calculate the fraction of correctly perceived intervals
  as a function of the distance between interval categories $\Imin$
  (Fig. \ref{fig:vmt_model}C). We then define an acceptable
  minimum interval size as that which corresponds to $99\%$ of intervals
  being correctly identified. This allows us to determine
  a minimum interval size which depends on $\sigmaprod$ and $\sigmaper$
  (Fig. \ref{fig:vmt_model}D).

  While we lack systematic studies of the range of 
  errors of humans' ability to sing or identify intervals accurately,
  a few studies provide information which allows one to speculate. 
  Studies on barbershop quartets and Indian classical music respectively report
  $\sigmaprod \approx 4-17$ cents and $\sigmaprod \approx 9-15$
  cents \cite{hagst80, seris11}.
  A study including untrained, amateur and professionals as subjects
  reported errors in sung intervals ranging from approximately $10$
  to $200$ cents \cite{pfojc17}. Studies on interval perception
  show that interval discrimination has a similar range
  of errors \cite{burja94, permp96}.
  One study reported just-noticeable differences ranging from $4$ cents
  (professional) to $\sim 40$ cents (untrained) \cite{mcdja10}. 
  Another study found that musicians who can correctly identify Western-classical interval categories
  accurate to $100$ cents generally did not discern inaccuracies
  of $20$ cents \cite{siepp77}. From this we suggest that expert musicians can be
  thought of having $\sigmaprod \approx \sigmaper \approx 10$ which
  results in a lower limit of $\Imin=54$. If minimising
  errors in transmission is important, it would be inadvisable
  to use scales with intervals close to the $\Imin$ of a
  trained expert. The predictions of this approach accord with the empirical
  observation that in reality such small intervals are rarely used.
  Arabic music is one of the few examples of a culture using ${\sim}50$ cents intervals, 
  and they are considered ornamental by many. For the untrained person,
  it is likely preferable that intervals should be considerably larger.

\subsection*{Vocal mistuning theory predicts equidistant scales}

  For any convex, monotonically decreasing cost function $C = f(I)$, 
  where $\sum\limits^{N}_{i} I_i = 1200$,
  $\sum\limits^{N}_{i} C_i$ will be minimized by setting each $I_i = 1200/N$.
  For a concave function, a minimum is found as one interval
  $j$ tends to the limit $I_j\to1200$ while the other intervals
  $k$ tend to zero $\sum\limits^{N-1} I_k \to 0$. 
  The marginal case is that of a linear function, which does
  not lead to a bias, implying that the theory
  does not affect the choice of scales.

  The vocal mistuning theory states that as interval size increases
  the probability of miscommunication decreases \cite{pfojc17}, but we do not know the
  form of the corresponding cost function. If we want to improve accuracy in
  transmission, are we more concerned with having fewer
  small intervals or having more large intervals?
  If it is the former, then the cost function is convex, 
  while if it is the latter then the function is concave. 
  Fig. \ref{fig:vmt_model}C shows that as $\Imin$
  increases the accuracy saturates. 
  This means that the cost function cannot diverge as $I$ increases, so
  it must be a convex function. Hence, the vocal mistuning theory
  predicts equidistant scales.

  Equidistant scales are typically considered rare \cite{savpn15}, and
  non-equidistant scales are professed to have advantages (e.g., tonality) over equidistant
  scales \cite{balcm80, kesmp84, schps96, mcdmp05}. However many scales are almost
  equidistant \cite{wacna50, vanam80, mcnja08, rosms17}, and there may be fewer
  equidistant scales simply because the base probability is low
  due to there being fewer ways to construct an equidistant scale.

  There remains the possibility that there is an optimal interval size,
  if one considers the effect of the vocal motor constraint theory, which
  states that larger intervals are harder to sing \cite{tiepn11}. By combining these two theories,
  of vocal mistuning and motor constraint, one may predict an optimal interval size.
  However we do not currently see a solid foundation upon which
  to construct such an analysis. One main issue is that singing and
  listening abilities vary widely across individuals, thus the optimal interval
  size would likely be best described by a broad distribution. Such
  a distribution is unlikely to be specific enough to account for the
  diversity of scales by itself, but it may be useful to consider this
  in tandem with other theories.

\subsection*{Model parameter sensitivity}

  In total, four tunable parameters are used in our generative models of scales:
  $\Imin$, $n$, $w$ and $\beta$.
  $\Imin$ is the minimum interval size allowed in a scale such that generated
  scales with any interval smaller than $\Imin$ are rejected.
  $n$ is a parameter in the cost function for the \TRANS~model;
  higher $n$ corresponds to larger deviations from a compressible
  interval template being penalized more heavily than smaller deviations.
  $w$ is the window size for the \HAR~and \FIF~model templates;
  higher $w$ corresponds to a greater tolerance for errors when perceiving
  consonant intervals. 
  $\beta$ controls the strength of any applied bias;
  the same value of $\beta$ in different models does not result in the
  same strength of the bias. We report instead $\log_{10} q$,
  where $q$ is the fraction of generated scales that are accepted by the model,
  such that increasing $\beta$ reduces $q$. $q$ is not a perfect indicator
  of bias strength, but it is easy to measure and scales with model
  performance better than $\beta$.
  The $q$ values for the different models (not reported in the main paper)
  are shown in Table \ref{tab:params}. For each model we generated $S=10^4$
  scales. Two metrics are used to characterize the performance of the models:
  $\dI$ and $\fD$. $\dI$ is the distance between the model and empirical $\IP$ distributions.
  $\fD$ is the fraction of scales from the database which are predicted by
  the model.

\begin{figure}
\centering
\includegraphics[width=\textwidth]{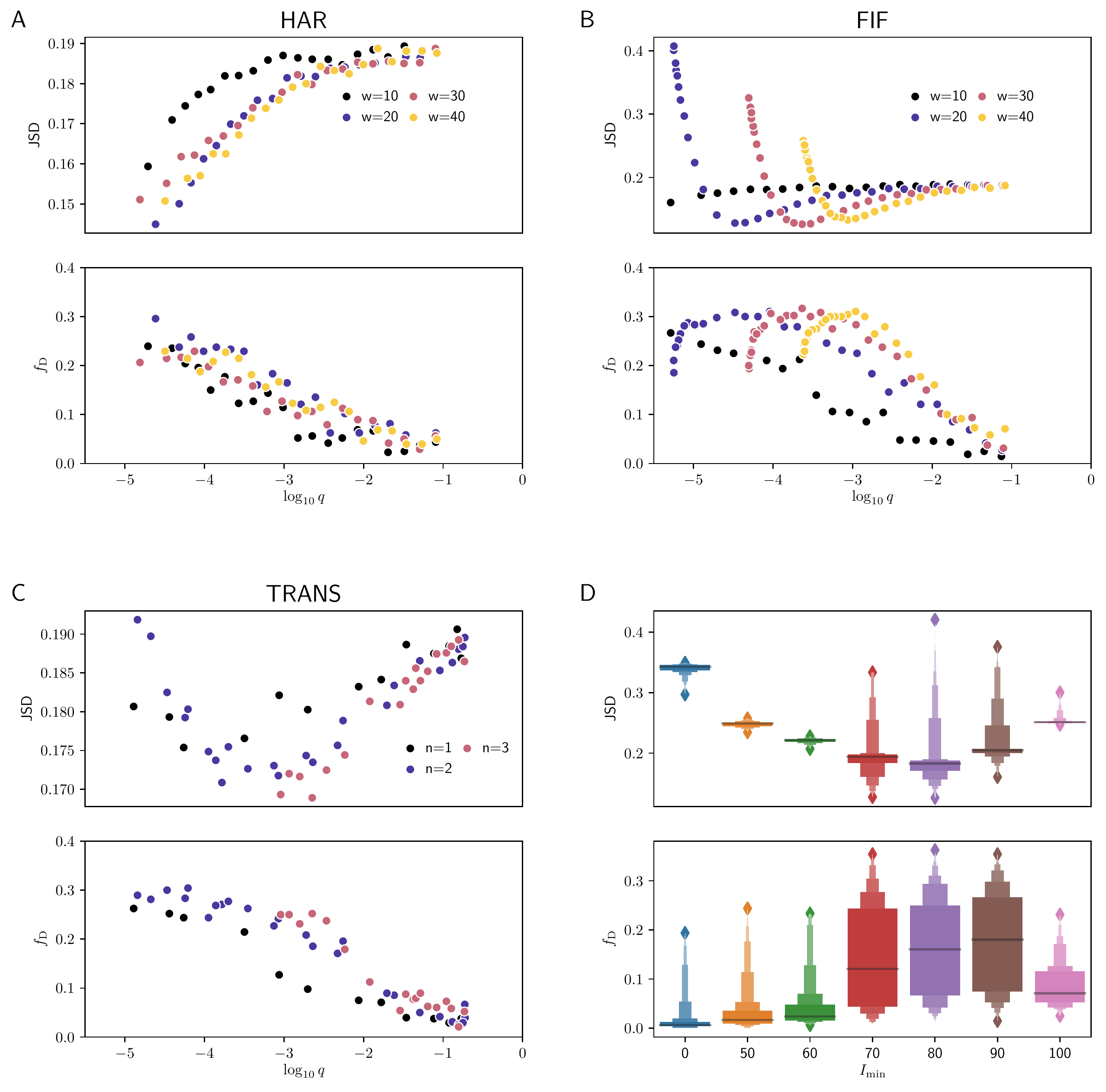}
\caption{\label{fig:sens}
  A-C: The Jensen-Shannon divergence between model and empirical distributions, $\textrm{JSD}$,
  and the fraction of real scales found by the models, $\fD$, 
  as a function of the fraction of generated scales, $q$, that are accepted by
  the model. This is reported for the three models (A) \HAR, (B) \FIF~ and (C) \TRANS,
  for $\Imin=80$.
  D: The distributions of $\dI$ and $\fD$ for all models
  as a function of the minimum interval size, $\Imin$. This includes
  results from all models.
  Note that the results for $\Imin = [70,80,90]$ appear qualitatively
  different to the results for other values of $\Imin$ --
  this is mainly due to differences in sampling.
  A-D: Results are only shown for $N=7$ for clarity.
}
\end{figure}

  For the \TRANS~model, increasing $n$ appears to improve the results,
  albeit slightly. For both the \HAR~and \FIF~models maximum performance
  is attained when $w=20-30$ (\ref{fig:sens}B-C). For all models shown there
  is a clear optimal $\log_{10} q$, which corresponds to an optimal $\beta$.
  For all the models, performance is greatest for $\Imin = 80-90$ cents (\ref{fig:sens}D).

  We chose to present results in Fig. \ref{fig:sens} for the \FIF~model
  using a different cost function than the one presented in the main text:
  \begin{equation}\label{eq:fif}
  C_{\FIF} = 1 - (N\bar F )^{2}~,
  \end{equation}
  where $\bar F$ is the fraction of intervals that are fifths, and $N$ is the number
  of notes in a scale. This is purely for purposes of illustration. When we use
  the cost function shown in the main text, the acceptance rate is too low
  to clearly show the optimal values of $\log_{10} q$, i.e., 
  the extrema in Fig. \ref{fig:sens}B would not be clear. In Eq. \ref{eq:fif}, $N$ is factor
  that normalizes the range of the cost function so that for all
  $N$, $0\leq C_{\FIF} \leq 1$. A full explanation
  of how the cost function affects the results is illustrated
  in Fig. \ref{fig:choo_a}, \ref{fig:choo_m} and \ref{fig:bias_sens}.

\begin{table}\centering
\caption{\label{tab:params}
  Full list of parameters, bias acceptance rate $q$ and Jensen-Shannon
  divergence ($\textrm{JSD}$) for each model presented in the main text Fig. 3, 4 and 6.
  We obtain the values for $\H_{min}$ and $\H_{max}$ from \MIN~model
  $\H$ distribution for each $N$ and $\Imin$. The $\textrm{JSD}$ is
  calculated as the distance between the $C_{\TRANS}$, $\H$ or $\bar F$ distributions
  obtained by the \MIN~model and the \TRANS, \HAR~or \FIF~model.
  }
\begin{tabular}{lccccccccc}
Model  &  $N$  & $I_{\textrm{min}}$ &  $n$  &  $w$  & $\H_{min}$ &  $\H_{max}$ & $\beta$ &  $q$  & $\textrm{JSD}$ \\ 
\toprule
RAN & 4 & 0 &  &  &  &  &    0.0 & 1.0e+00 &  \\
RAN & 5 & 0 &  &  &  &  &    0.0 & 1.0e+00 &  \\
RAN & 6 & 0 &  &  &  &  &    0.0 & 1.0e+00 &  \\
RAN & 7 & 0 &  &  &  &  &    0.0 & 1.0e+00 &  \\
RAN & 8 & 0 &  &  &  &  &    0.0 & 1.0e+00 &  \\
RAN & 9 & 0 &  &  &  &  &    0.0 & 1.0e+00 &  \\
MIN & 4 & 80 &  &  &  &  &    0.0 & 7.9e-01 &  \\
MIN & 5 & 80 &  &  &  &  &    0.0 & 5.9e-01 &  \\
MIN & 6 & 80 &  &  &  &  &    0.0 & 3.7e-01 &  \\
MIN & 7 & 80 &  &  &  &  &    0.0 & 2.0e-01 &  \\
MIN & 8 & 80 &  &  &  &  &    0.0 & 8.4e-02 &  \\
MIN & 9 & 80 &  &  &  &  &    0.0 & 2.5e-02 &  \\
HAR & 4 & 80 &  & 20 & 14.0 & 43.98 &    3.0 & 5.6e-02 & 0.12 \\
HAR & 5 & 80 &  & 20 & 15.0 & 41.67 &    7.0 & 2.3e-03 & 0.38 \\
HAR & 6 & 80 &  & 20 & 16.0 & 39.47 &   13.0 & 2.9e-05 & 0.66 \\
HAR & 7 & 80 &  & 20 & 17.0 & 37.57 &    9.5 & 1.7e-04 & 0.51 \\
HAR & 8 & 80 &  & 20 & 18.0 & 35.58 &    9.0 & 6.4e-05 & 0.42 \\
HAR & 9 & 80 &  & 20 & 18.0 & 31.84 &   14.0 & 2.6e-06 & 0.71 \\
TRANS & 4 & 80 & 2 &  &  &  &  200.0 & 7.8e-02 & 0.48 \\
TRANS & 5 & 80 & 2 &  &  &  &  284.8 & 1.3e-02 & 0.62 \\
TRANS & 6 & 80 & 2 &  &  &  & 1666.7 & 3.9e-05 & 0.82 \\
TRANS & 7 & 80 & 2 &  &  &  &  471.4 & 2.0e-04 & 0.77 \\
TRANS & 8 & 80 & 2 &  &  &  &  412.5 & 6.2e-05 & 0.76 \\
TRANS & 9 & 80 & 2 &  &  &  &  500.0 & 6.0e-06 & 0.79 \\
FIF & 4 & 80 &  & 20 &  &  & 2000.0 & 3.2e-03 & 0.60 \\
FIF & 5 & 80 &  & 20 &  &  & 2000.0 & 1.2e-03 & 0.61 \\
FIF & 6 & 80 &  & 20 &  &  & 4000.0 & 3.9e-06 & 0.71 \\
FIF & 7 & 80 &  & 20 &  &  & 4000.0 & 9.4e-07 & 0.72 \\
FIF & 8 & 80 &  & 20 &  &  & 4000.0 & 9.7e-08 & 0.72 \\
FIF & 9 & 80 &  & 20 &  &  & 4000.0 & 9.1e-09 & 0.73 \\
\end{tabular}
\end{table}

\subsection*{Packing of fifths produces interval mixing rules for scales}

\begin{figure}
\centering
\includegraphics[width=\textwidth]{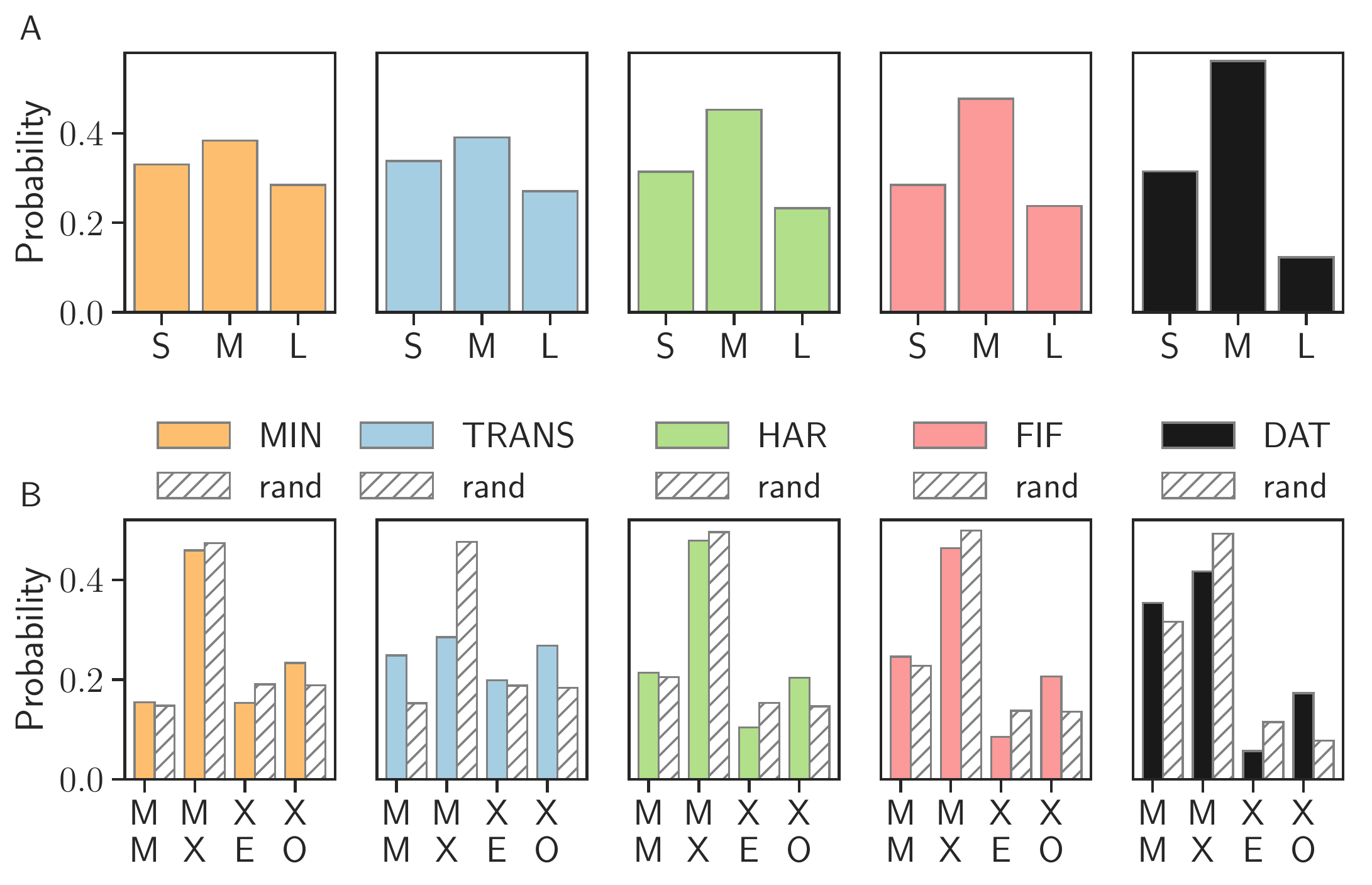}
\caption{\label{fig:mix1}
  A: The distribution of interval sizes (\textbf{S}: small, \textbf{M}: medium, \textbf{L}: large)
     in populations for \MIN, \TRANS, \HAR, \FIF, and \DAT.
  B: The probability that interval sizes are found adjacent to each other
     in populations, compared with the probabilities of random pairing;
     \textbf{M}-\textbf{M} (medium interval beside a medium interval), \textbf{M}-\textbf{X}
     (\textbf{X}: extreme, which includes \textbf{S} and \textbf{L}),
     \textbf{X}-\textbf{E} (\textbf{E}: equivalent, which includes
     \textbf{S}-\textbf{S} or \textbf{L}-\textbf{L}),
     \textbf{X}-\textbf{O} (\textbf{O}: opposite, which includes
      \textbf{S}-\textbf{L} and \textbf{L}-\textbf{S}).
     We calculate probabilities without considering the order in which
     the intervals are chosen.
  A-B: Results are only shown for $N=7$ for clarity.
}
\end{figure}

  Our generative model arranges intervals randomly into scales.
  To see if scales from the database are arranged randomly,
  we calculated the fraction of small (\textbf{S}: $\IP < (1-x)1200/N$),
  medium (\textbf{M}: $(1-x)1200/N \leq \IP \leq (1+x)1200/N$) and
  large intervals (\textbf{L}: $\IP > (1+x)1200/N$), using $x=0.2$. We use
  these formulae to define size categories since we are
  interested in relative sizes rather than absolute sizes.
  Using this fractions, we calculated the probability that
  certain intervals are found adjacent to each other by mixing randomly,
  and compared these results with the \DAT~and model-generated
  scales (Fig. \ref{fig:mix1}). We find that the \DAT~scales are much more likely to
  have small intervals placed with large intervals and vice versa
  than random mixing would predict.

\begin{figure}
\centering
\includegraphics[width=0.6\textwidth]{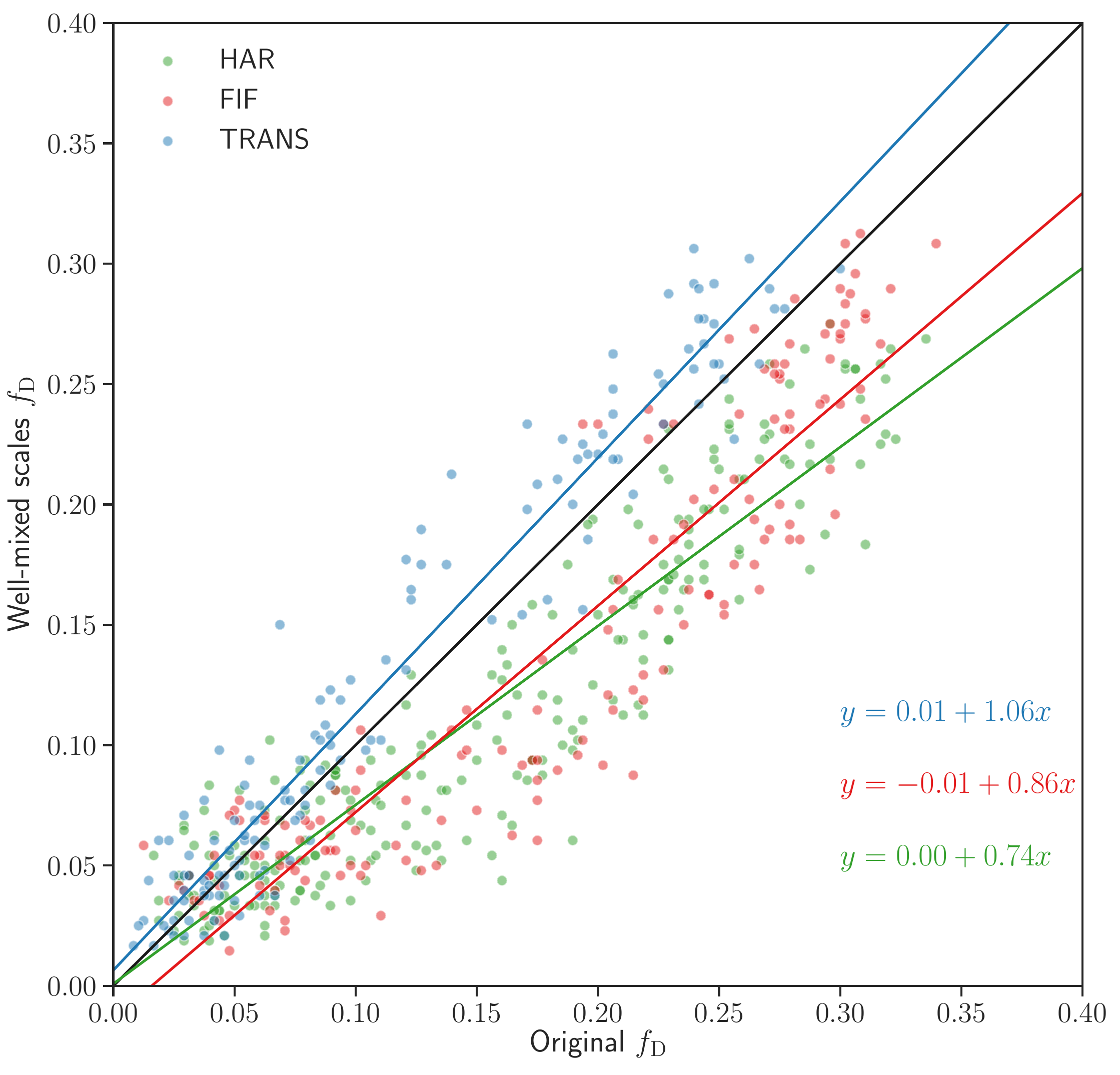}
\caption{\label{fig:mix2}
  The fraction of real scales $\fD$ originally found by the models
  plotted against $\fD$ calculated after scales are rearranged
  with a bias towards being well-mixed; results are shown for the \HAR,
  \FIF~ and \TRANS~ models.
}
\end{figure}

  To investigate the influence of the above effect, 
  we rearranged the intervals in our model-generated scales by
  biasing them so that they are well-mixed. Scales are considered well-mixed
  if the sum of two consecutive intervals is close to the average sum
  of two intervals, $2\times 1200/N$. This means that small intervals are placed
  beside large intervals so that their combined size approximates
  that of two medium intervals. To arrange scales so that they are mixed,
  we calculate the cost function $\Cmixj$ for all unique permutations, $M$,
  of a set of intervals as
  \begin{equation}
  \Cmixj = \left[ \frac{1}{N} \sum^{N}_{i} 
  \left(I_i + I_{i+1} - \frac{2400}{N} \right)^2 \right]^{\frac{1}{2}}~,
  \end{equation}
  where $N$ is the number of notes in a scale, $j$ is the scale index,
  and $I_i$ is the $i^{th}$ pair interval in a scale. 
  When the subscript $i>N$, due to the circular nature of scales, $i \to i - N$.
  We normalize $\Cmixj$ to get  $\tilde{C}_{\mathrm{mix},j}$ for each $j$ 
  by dividing by the maximal $\Cmixj$.
  We then randomly draw a scale, with the probability
  of a scale $j$ being picked, $P_{\mathrm{mix},j}$,
  \begin{equation}
  P_{\mathrm{mix},j} = \frac{\exp{\tilde{C}_{\mathrm{mix},j}^{-1}}} 
  {\sum^{M}_{k=1} \exp{\tilde{C}_{\mathrm{mix},k}^{-1}}}.
  \end{equation}
  Arranging scales in this way results in higher $\fD$ values for
  the \TRANS~model, but lower $\fD$ values for the \HAR~and \FIF~models
  (Fig. \ref{fig:mix2}). Thus, mixing scales in
  this way improves the results for the \TRANS~model since it
  does not take interval order into account in its bias. However for  
  the harmonicity models, which do take interval order into account,
  this method of ordering intervals results in an inferior fit.
  This indicates that scales are in general well-mixed rather than random,
  but it is more important that they are ordered such that they
  maximise the number of fifths, and perhaps to an extent other harmonic intervals.

\subsection*{Correlations between harmonicity models}

\begin{figure}
\centering
\includegraphics[width=0.8\textwidth]{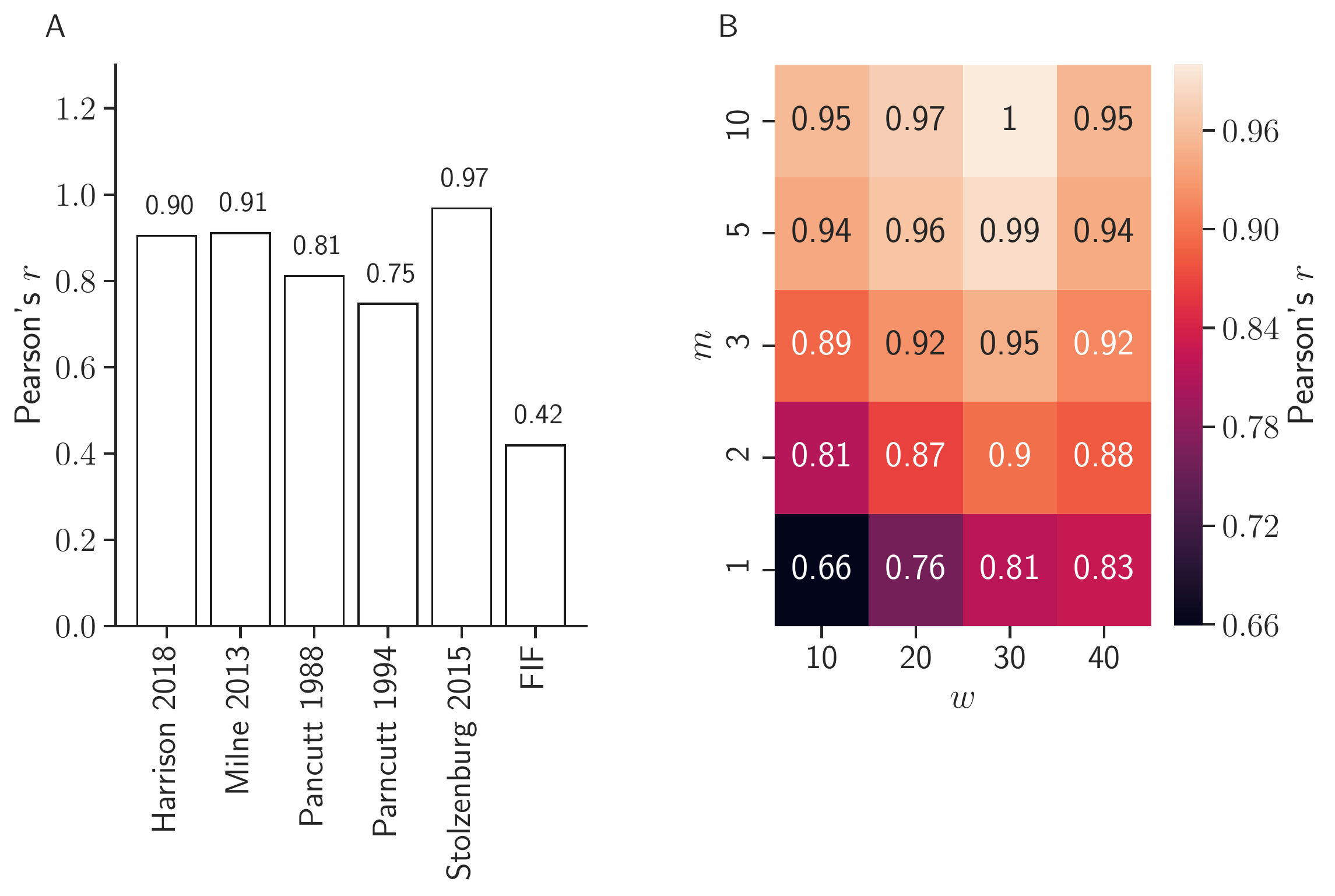}
\caption{\label{fig:har_corr}
  A: Correlations between the model of Gill and Purves (2009) and six other
  harmonicity models.
  B: Correlations between the average harmonicity score, $\H$, and
  the fraction of fifths, $\bar F$, for a population of scales (\MIN~model,
  $N=7$, $S=10^4$). For each comparison, both models use the same value
  of $w$.
}
\end{figure}

  There are multiple models of harmonicity, which 
  primarily differ in how they deal with deviations
  from exact integer ratios and how they weight different harmonics.
  Using the \textit{incon} package \cite{harpr19} in R we tested five models
  \cite{parmp88,parpn94,mil13,frijm15,harar18} for
  correlations with the model of Gill and Purves (2009)
  (\HAR~model) by comparing their predictions
  for the twelve diatonic intervals. Pearson's $r$ values of $r>=0.75$ indicate
  that the models are significantly correlated (Fig. \ref{fig:har_corr}A). We compared
  the \HAR~and \FIF~models in the same way and found that they are less
  correlated (Fig. \ref{fig:har_corr}A).
  For the purposes of this study, however, it is more appropriate to
  check correlations between the average scores of scales rather than intervals.
  For a set of scales (\MIN~model, $N=7$, $S=10^4$) we calculate for each
  scale the average harmonicity score as 
  \begin{equation} \label{eq:s_har}
  \H = \left[ \sum^{N-1}_{i=0} \sum^{i+N-1}_{j=i+1} H(I_{ij}, w)^m/100^{m-1} \right]
           \bigg[N(N-1)\bigg]^{-1}~,
  \end{equation}
  \noindent where $H(I)$ is the harmonic similarity score 
  of an interval size $I$, $w$ is the size of the window 
  in which intervals are considered equivalent,
  and $I_{ij}$ is the interval between note $i$ and note $j$.
  The index $i=0$ refers to the starting note of the scale 
  and $j$ takes into account the circular nature of scales
  (if $j>N$ then note $j$ is an octave higher than note $j-N$).
  The $100^{m-1}$ term is a normalization factor to ensure
  that $S_{\HAR}\leq100$.
  Note that in the \HAR~model presented in the main text  we use $m=1$,
  while $m=2$ and $m=3$ correspond to \HAR$^2$ and \HAR$^3$ in the main text Fig. 5.
  The harmonic similarity score of an interval is calculated 
  from its frequency ratio expressed as a fraction.
  \begin{equation}
  H(I) = \frac{x+y-1}{xy}\times100~,
  \end{equation}
  \noindent where $x$ is the numerator and $y$ is the denominator of the fraction.
  The harmonic similarity template is produced by creating a grid 
  of windows of maximum size $w$ centred about the intervals 
  that have the largest $H$ values. An interval expressed in cents
  is allocated to the window with the highest $H$ value that is within $w/2$ cents.
  For the same set of scales we calculate the fraction of fifths as
  \begin{equation}
  \bar F = \left[ \sum^{N-1}_{i=0} \sum^{i+N-1}_{j=1} \FIF(I_{ij}, w) \right]
           \bigg[N(N-1)]\bigg]^{-1}~,
  \end{equation}
  \noindent where $\FIF(I)=1$ if $|I-702| \leq w/2$. 
  We show that for if $w>=10$ there is a strong correlation
  between average \HAR~and \FIF~scores for scales (Fig. \ref{fig:har_corr}B).
  This correlation increases further as $m$ increases. As 
  $m\to\infty$ the \HAR~model becomes a linear function of the \FIF~model.

\subsection*{Scales that are found, and not found by the models}

\begin{figure}
\centering
\includegraphics[width=\textwidth]{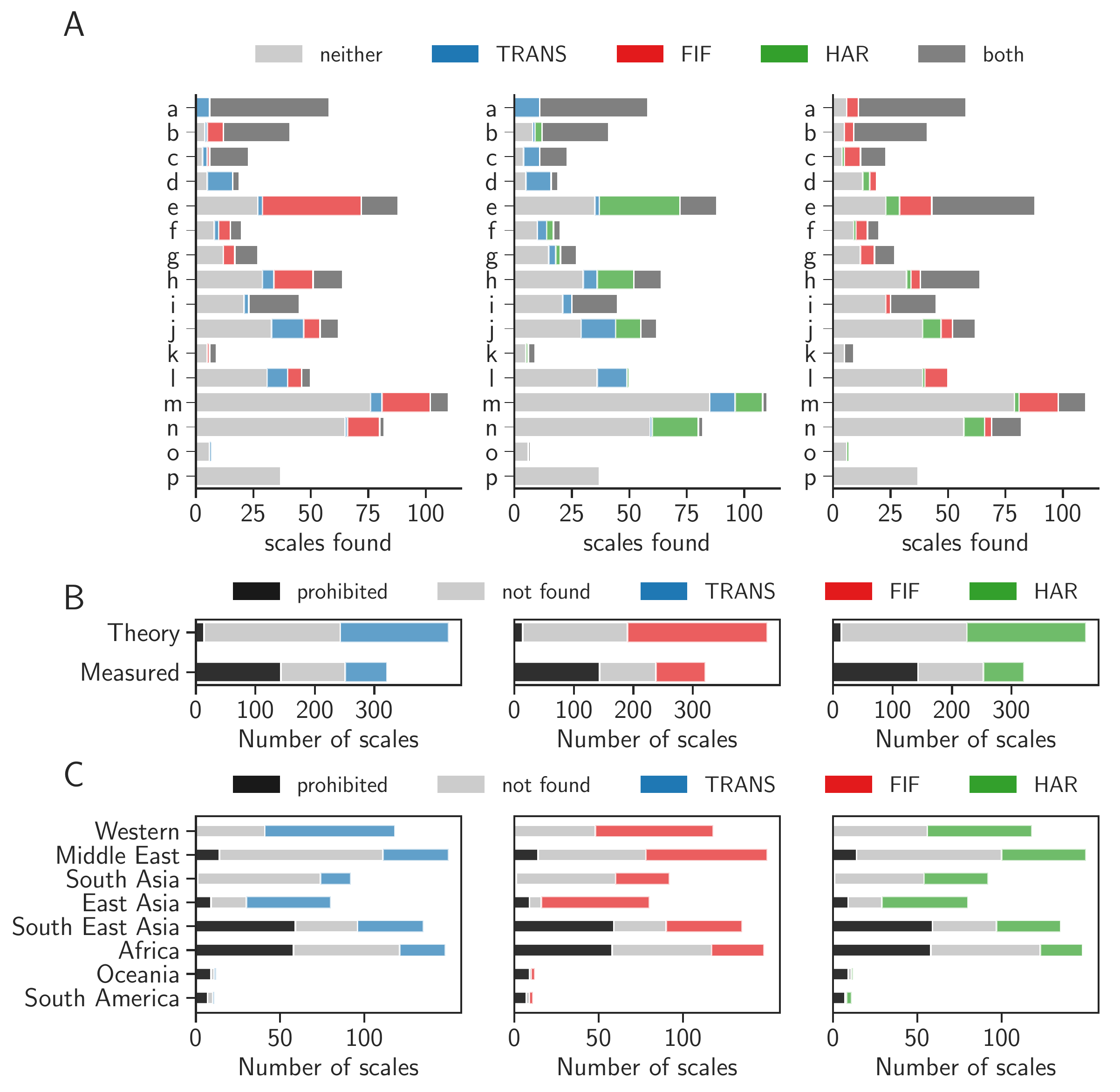}
\caption{\label{fig:found_scales}
  A: Comparison of the scales found by two models for three sets:
  \TRANS~and \FIF; \TRANS~and \HAR; \HAR~and \FIF. Stacked bars indicate
  what scales are found by: neither model; only one of either model;
  both models.
  B: Number of `theory'/`measured' scales that are not found,
  `prohibited' (it is not possible to find them
  due to the hard constraints of the model), and found by each model.
  C: Number of scales from each continent that are not found,
  `prohibited' and found by each model.
}
\end{figure}

  We show additional counts of scales that are found or not
  found by the three models across three categories (Fig. \ref{fig:found_scales}):
  16 clusters that are based on the scales' adjacent interval sets;
  scale type (`theory' or `measured'); geographical region (continent).
  The fact that the models do not particularly depend on scale type
  or geographical region are good indicators that the models
  capture properties of scales that are general. In particular
  we note that only scales from the `theory' type are known
  to have been at one point explicitly based on fifths.

\subsection*{Correlations between transmittability and harmonicity models}

\begin{figure}
\centering
\includegraphics[width=0.8\textwidth]{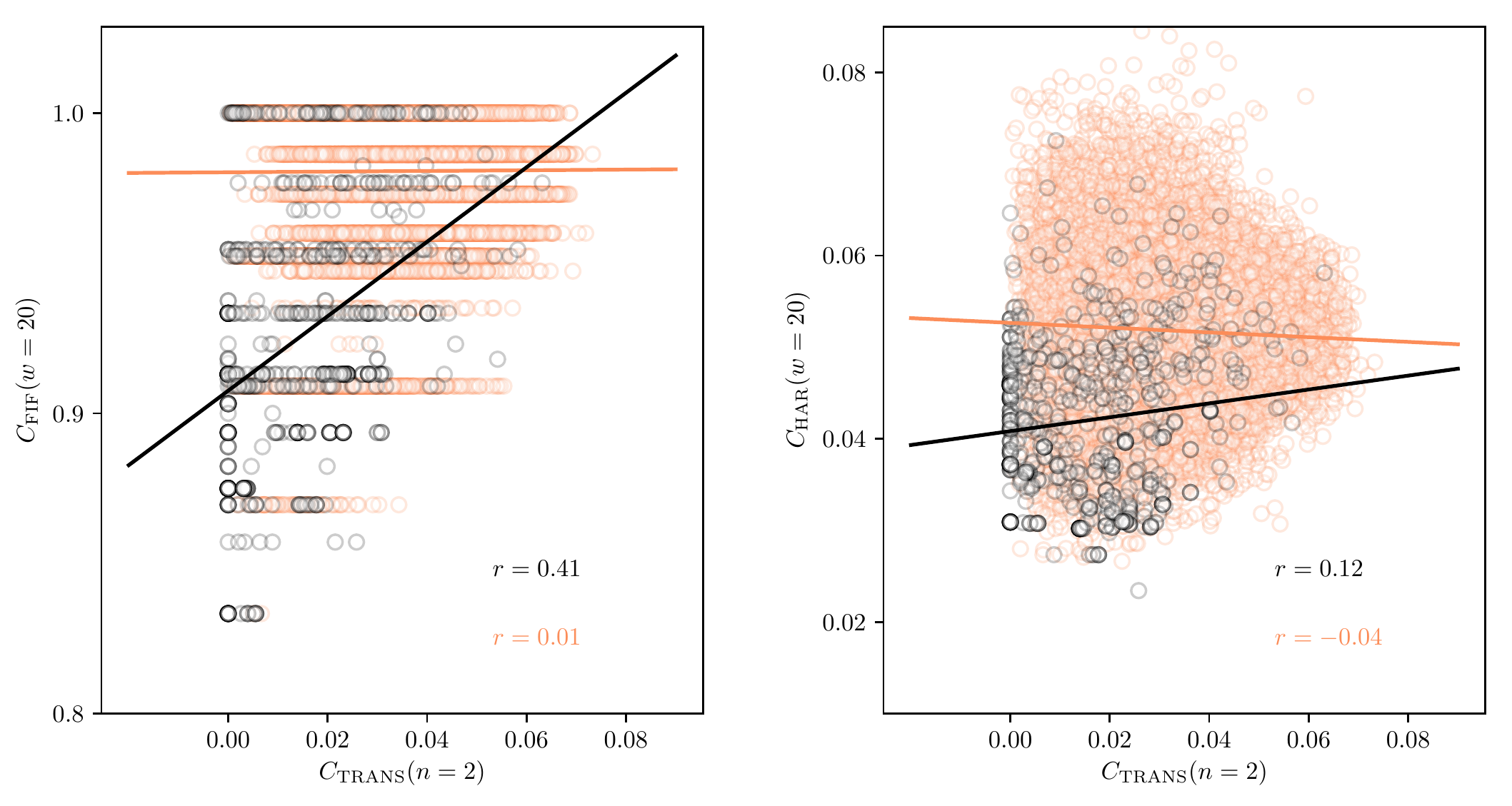}
\caption{\label{fig:trans_fif_corr}
  Correlations between transmittability and harmonicity cost functions for two sets of
  data: \DAT~scales ($S=742$, black) and \MIN~scales ($S=10^4$, orange).
  Pearson's $r$ is indicated in the plots for each correlation.
}
\end{figure}

  We tested for correlations between the transmittability and harmonicity
  models by calculating the correlations between cost functions in two
  populations of scales: \DAT~scales ($S=742$) and \MIN~scales ($S=10^4$). The results
  for the \MIN~scales indicate that in general the models
  are not correlated (Fig. \ref{fig:trans_fif_corr}). However, after scales have been selected by
  humans (\DAT~scales) there is a significant ($p<10^{-30}$) correlation
  between the \TRANS~and \FIF~models, while there is a weak but significant ($p<10^{-3}$)
  correlation between the \TRANS~and \HAR~models.

\subsection*{Tritone intervals are scarce due to packing of fifths}

\begin{figure}
\centering
\includegraphics[width=0.6\textwidth]{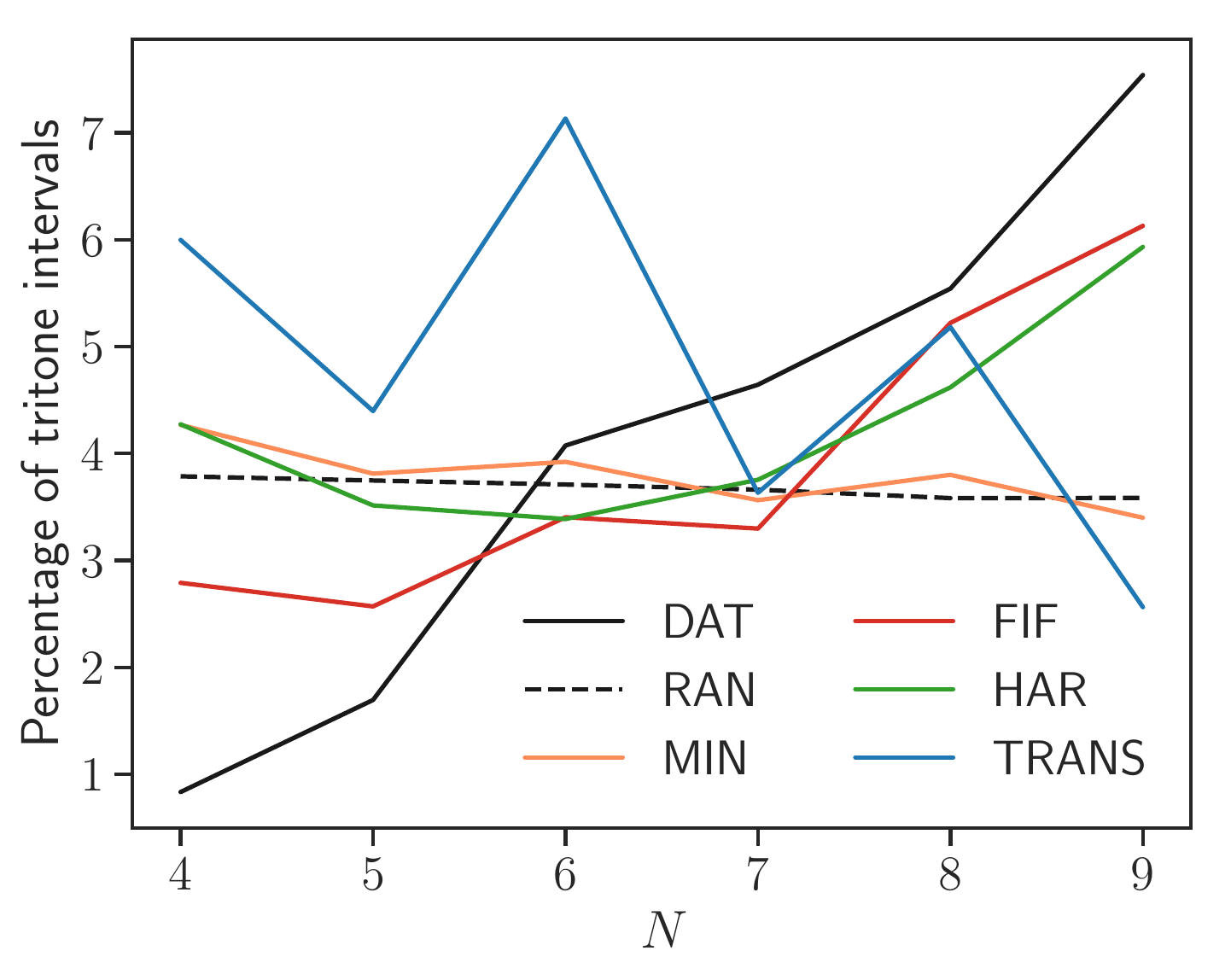}
\caption{\label{fig:tritone}
  The percentage of tritone intervals found in $N$ note scales
  in our database and in scales generated by the models reported in Fig. 3.
}
\end{figure}

  The frequency of tritone intervals in our database, $f_t$, is calculated as a
  function of $N$ such that
  \begin{equation}
  f_t = \frac{1}{N(N-1)S_N} \sum_{i}^{S_N} t_{i}
  \end{equation}
  where $N$ is the number of notes in a scale,
  $t_{i}$ is the total number of tritone intervals in
  scale $i$, and $S_N$ is the sample size of $N$ note scales.
  We consider any interval which is $600 \pm 20$ cents to be
  a tritone, given that accepted tritone frequency ratios $10/7$ and $7/5$
  correspond to $618$ and $583$ cents respectively.
  The database (\DAT) scales shows that the fraction
  of tritone intervals increases linearly with $N$ (Fig. \ref{fig:tritone}).
  The model that best reproduces this trend is the \FIF~model.

\begin{figure}
\centering
\includegraphics[width=0.80\textwidth]{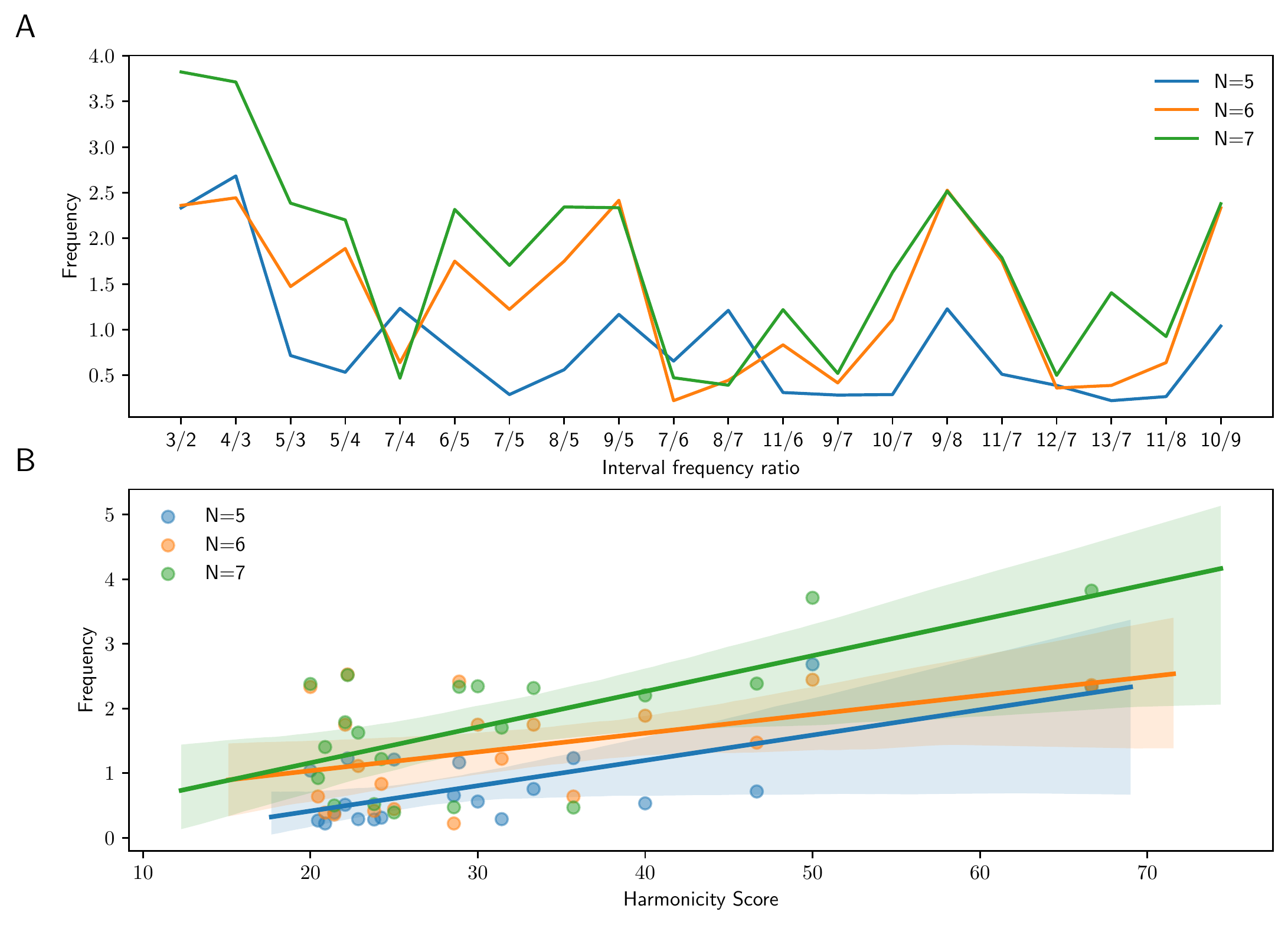}
\caption{\label{fig:harm_int}
  A: The frequency of intervals found in \DAT~scales, where the intervals
  are ordered according to their harmonicity score.
  B: Frequency of intervals found in \DAT~scales, as a function of harmonicity.
  Regression lines are shown with shaded regions indicating $95\%$ confidence
  intervals ($N=5$, $r=0.70$, $p<0.005$; $N=6$, $r=0.43$, $p>0.05$;
  $N=7$, $r=0.65$, $p<0.005$; SI Fig. 9).
}
\end{figure}

\subsection*{Thai and Gamelan scales are variable individually, but coherent as an ensemble}

\begin{figure}
\centering
\includegraphics[width=0.80\textwidth]{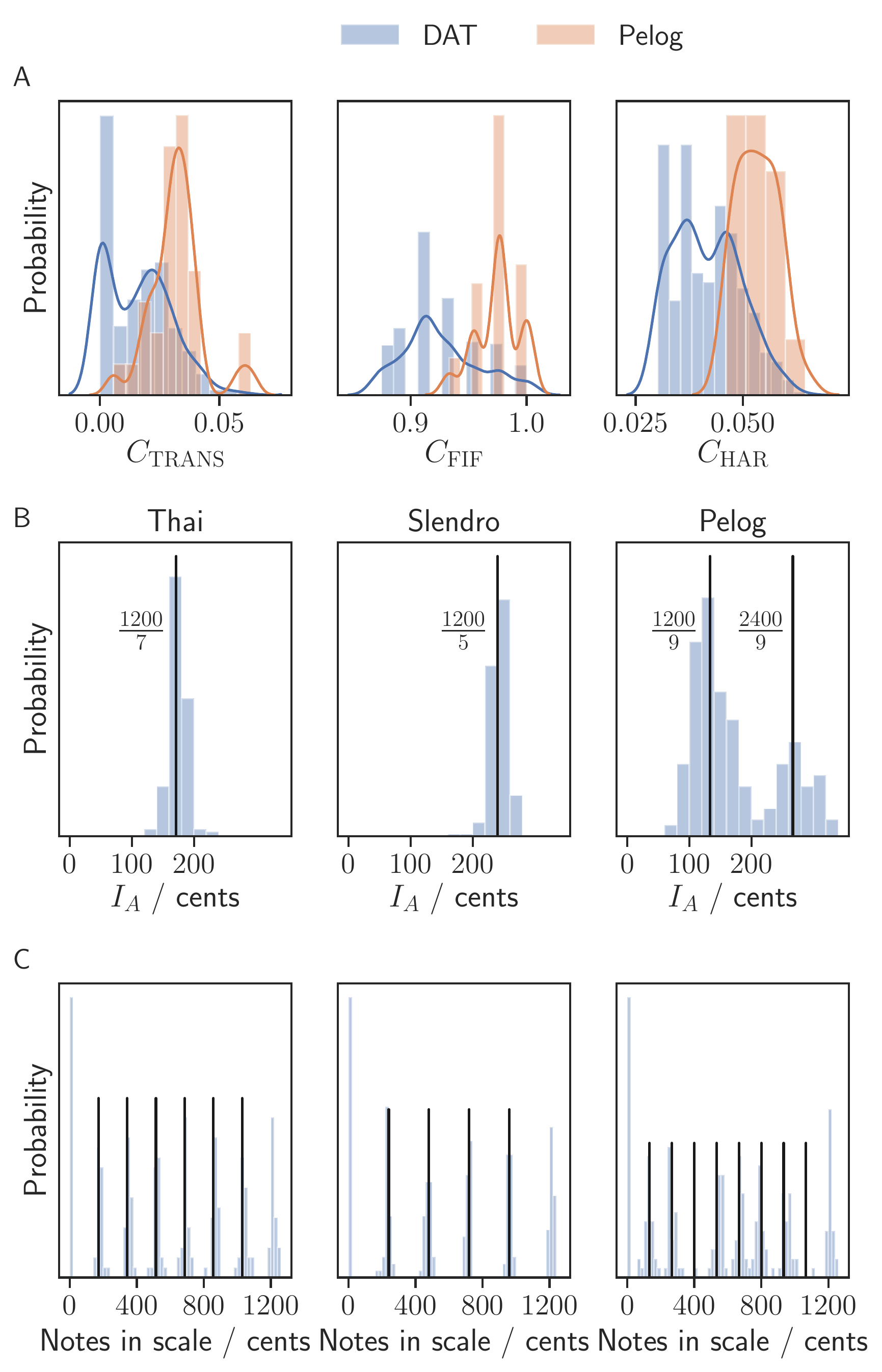}
\caption{\label{fig:thai}
  A: The distributions of the cost functions for pelog scales (orange)
  and all \DAT scales (blue: including the pelog scales).
  B-C: The probability distributions of pair intervals, $\IP$, (B) and scale
  notes (C) for three types of scales: the 7 note Thai tuning, the 
  5 note Gamelan slendro scale, and the 7 note Gamelan pelog scale.
  Lines correspond to exactly equidistant scales for 7-tet, 5-tet and
  9-tet tunings.
}
\end{figure}

  Pelog scales are unlikely to be predicted by any model since they have high costs
   compared to other scales (Fig. \ref{fig:thai}A). The \TRANS~model
  fits pelog scales better than the other models.

  The Thai tuning is considered to be equidistant, and intervals from
  an ensemble of Thai tunings can indeed be approximated by a Gaussian
  distribution with a mean $\mu=1200/7$ (Fig. \ref{fig:thai}B).
  However, intervals within individual scales can deviate
  wildly from this theoretical ideal, with ranges of up to $96$ cents
  observed in our database. The tunings of Gamelan slendro and pelog scales
  exhibit similar behaviour (Fig. \ref{fig:thai}C).

\subsection*{Results are robust to sub-sampling of the database}

\begin{figure}
\centering
\includegraphics[width=0.9\textwidth]{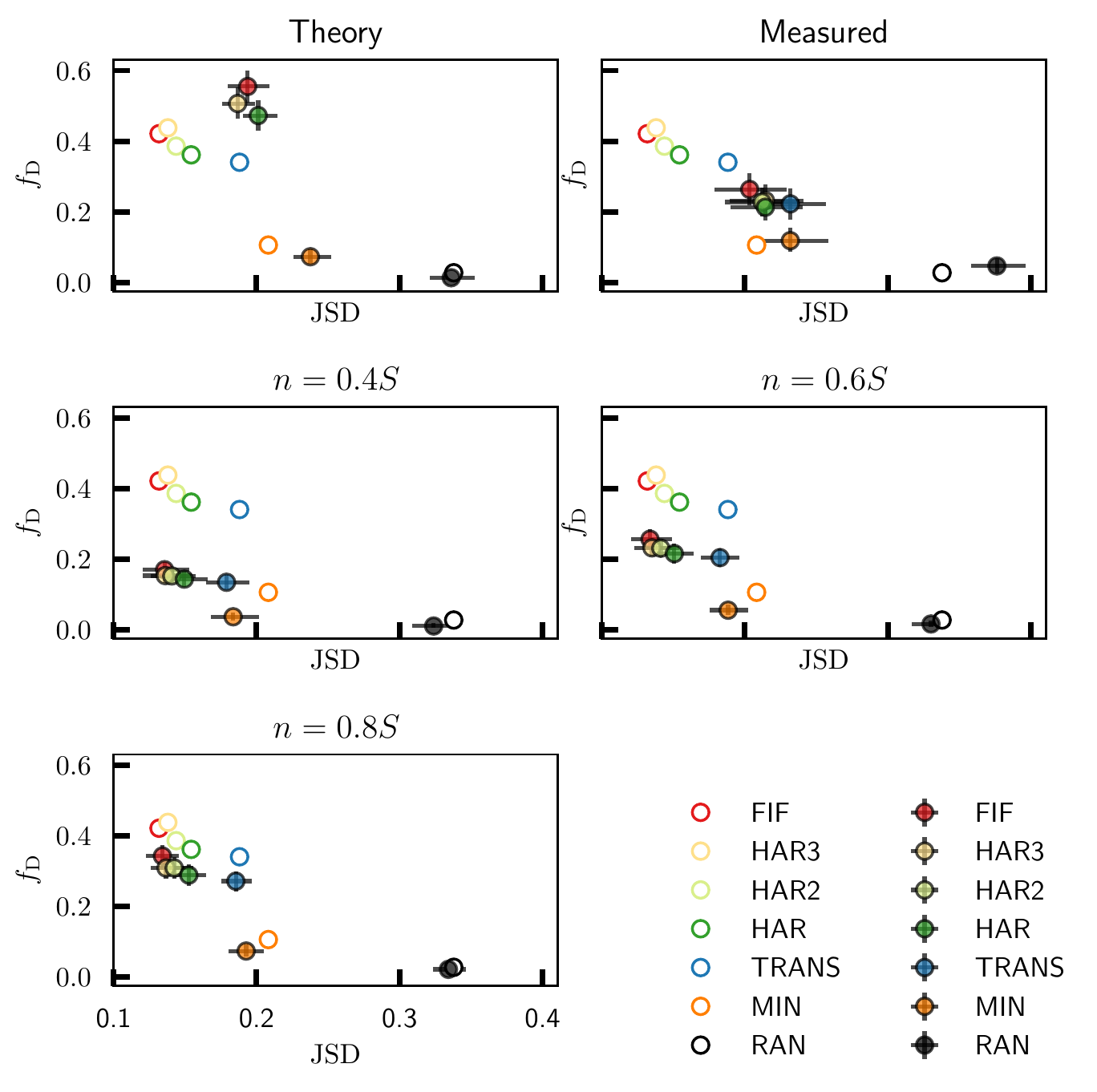}
\caption{\label{fig:data_sens}
  Results for models tested against resampled data: results are shown
  for Theory (only the `theory' scales), Measured (only the `measured' scales)
  and three sizes of bootstrapped samples. Whiskers indicate $95\%$ confidence
  intervals. Results for the best performing models using the full data
  set, \DAT, are shown in each plot as empty circles.
}
\end{figure}

  We repeat our analysis for the best performing models for each model
  shown in the main text Fig. 5 on sub-samples of the database.
  We use three types of sub-samples: all of the `theory' scales;
  all of the `measured' scales; bootstrapped resamples. We created
  $10$ bootstrapped resamples for each size: $n=0.4S$, $n=0.6S$ and $n=0.8S$;
  where $S$ is the size of the sample from which the resamples
  are drawn. We only used data for $N=5$ and $N=7$ due to the small
  sample sizes of other $N$. Comparing the results from main text Fig. 5
  to the results of the sub-samples we found that under most cases the
  results do not qualitatively differ (Fig. \ref{fig:data_sens}).
  The main exception is that for the sub-sample which only includes `measured'
  scales, the harmonicity models have relatively low values of $\dI$.
  This indicates that when only considering `measured' scales, the
  harmonicity models do not reproduce the adjacent interval
  distributions more accurately than chance (\MIN~model).

\subsection*{Hexatonic scales are not so rare as the database suggests}

\begin{figure}
\centering
\includegraphics[width=0.8\textwidth]{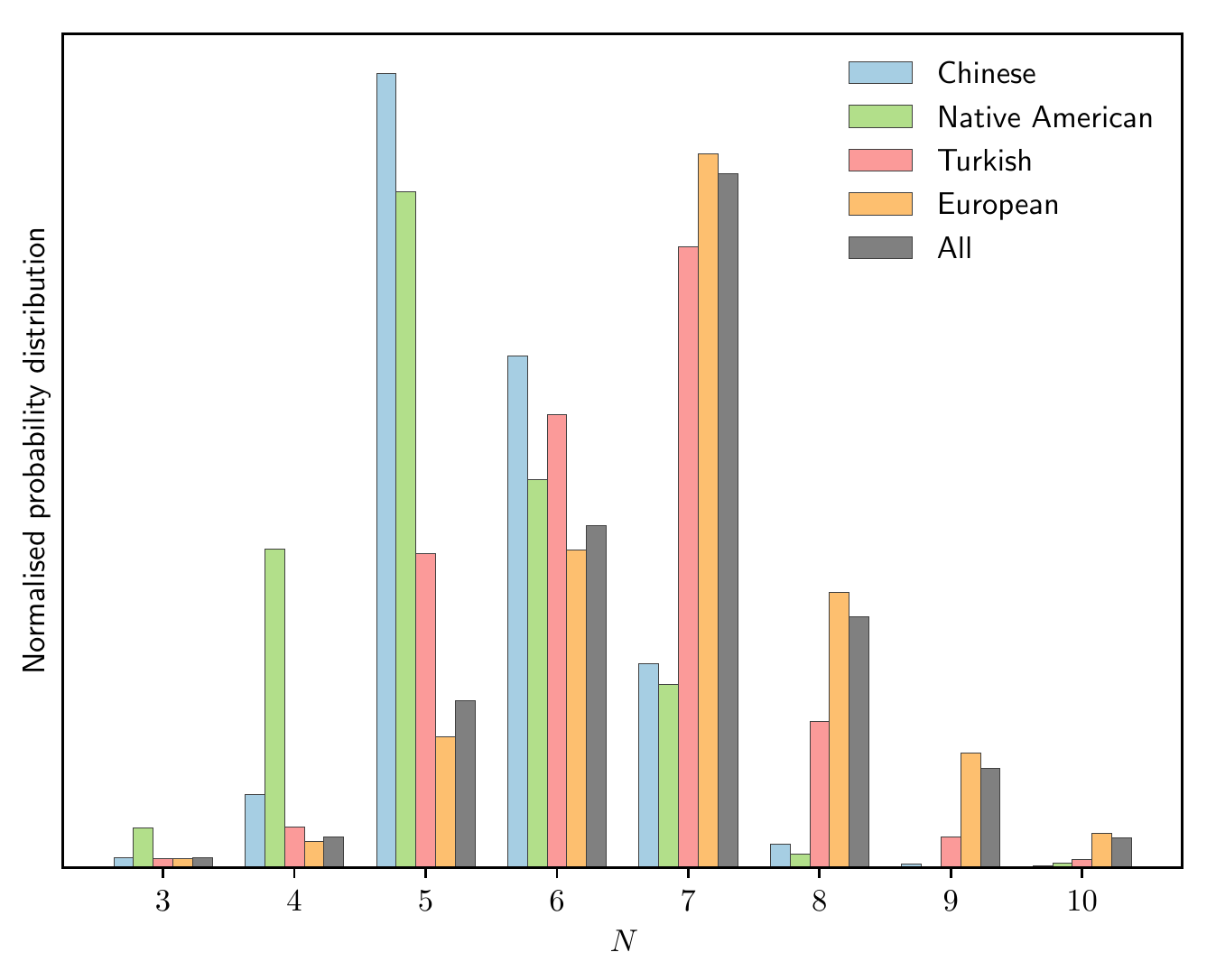}
\caption{\label{fig:essen}
  The probability distributions of number of notes $N$ in folk songs
  from different cultures (Chinese, Native American, Turkish, European).
}
\end{figure}

  Using several databases, we studied the distribution of $N$ unique notes
  used in folk melodies across cultures. The databases include:
  Essen folk song collection (Chinese and European) \cite{sch95};
  KernScores humdrum database (Native American, Polish, European) \cite{sapis05};
  the Meertens tune collection (Dutch) \cite{van19};
  Uzan Hava humdrum database (Turkish) \cite{sen11}.
  While all cultures shown here have a preference for
  either 5 or 7 notes in their songs, 6 note
  songs are consistently the second most frequent (Fig. \ref{fig:essen}).
  This effectively means that six note scales are actually quite prevalent, 
  despite the fact that they are rarely counted as scales.

\subsection*{Effect of the functional form of the cost function}

  Since we are using a Boltzmann distribution
  \begin{equation}
    P = \min\{1, \exp(-\beta C)\}~,
  \end{equation}
  \noindent for each of the three models we need a cost function which
  approaches zero for the type of scales that the model promotes. In the case
  of the \TRANS~model the cost function already satisfies this condition --
  $C_{\TRANS}$ is a measure of the deviations from a compressible template, such
  that a scale with $C_{\TRANS}=0$ maximises lossless compression.
  For the \HAR~and \FIF~models, however, we want to maximise respectively
  the average harmonic score $\H$ and the fraction of fifths $\bar F$.
  For purposes of illustration, in the following we consider how
  to choose a cost function for the \HAR~model.

  The exact functional form of the cost function is not necessarily important.
  We are interested instead in how it affects the results of the model.
  There are two main effects of the cost function: the acceptance rate
  and the selectivity. The acceptance rate is simple to measure,
  \begin{equation}
    \textrm{acc} = \sum P_{\textrm{acc}}(\H)~,
  \end{equation}
  \noindent where $P_{\textrm{acc}}(\H)$ is the probability that a scale with
  $\H$ is accepted. One measure for selectivity is the
  relative entropy between generated and accepted $\H$ distributions.
  However two different sets of distributions can give the same relative entropy.
  In particular, non-linear changes to the cost function will result in a different
  $\H$ distribution for the same relative entropy (linear changes are cancelled out
  by $\beta$ which is varied independently). Thus there is no obvious optimal
  way of constructing a cost function.

\begin{figure}
\centering
\includegraphics[width=\textwidth]{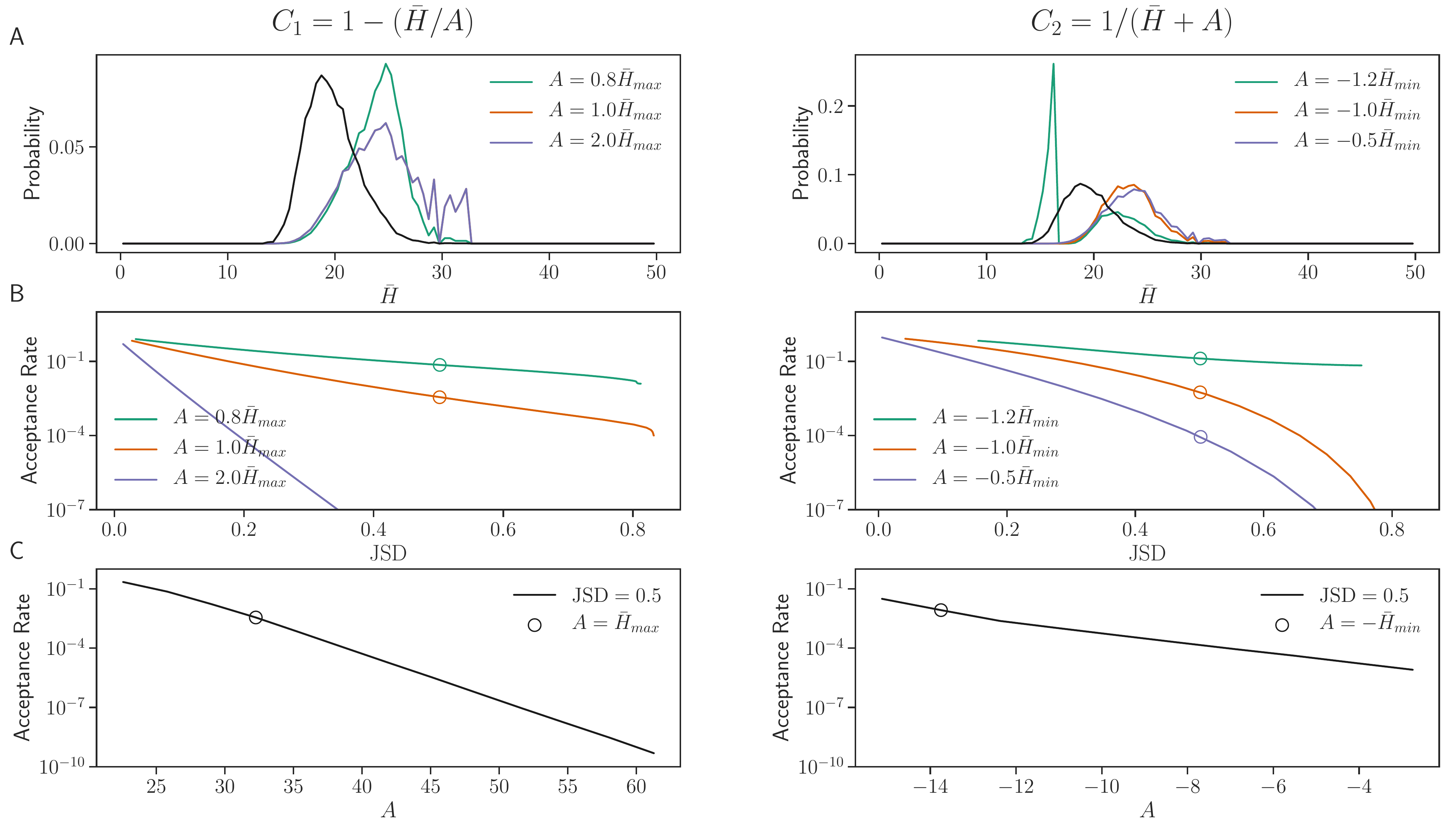}
\caption{\label{fig:choo_a}
  Effect of changing $A$ in the cost functions $C_1$ (left) and $C_2$ (right).
  A: Probability distribution of $\H$ for a population of $N=7$ scales
  accepted by the \MIN~model (black line). Theoretical probability
  distributions of $\H$ accepted by models with the cost functions
  $C_1$ (left) and $C_2$ (right). Values of $A$ correspond to:
  $A$ chosen so that some scales are always accepted (green line),
  $A$ chosen so that the acceptance is maximized, and $C>=0$ (orange line),
  $A$ chosen so that the acceptance is sub-optimal and $C>=0$ (purple line).
  $\beta$ is chosen so that the selectivity (JSD) is $0.5$.
  B: Acceptance rate vs. selectivity (JSD) for the same values of $A$ as in A.
  Increasing $\beta$ decreases acceptance rate, and increases selectivity (JSD).
  Circles correspond to a selectivity (JSD) of $0.5$.
  C: Dependence of the acceptance rate on $A$. Selectivity is constant at $0.5$.
  Circles correspond to the points where $A$ maximizes selectivity (JSD)
  while maintaining $C>=0$.
   }
\end{figure}

  Our approach is to design a cost function which allows us to optimize the bias strength
  via $\beta$ for the best model performance. We want
  to maximize the acceptance rate, while maximizing the selectivity, so that
  the model can reach the optimum selectivity in reasonable time.
  We study the effect of the functional form analytically, given a
  probability distribution of $\H$. In the following we use the
  $\H$ distribution resulting from the \MIN~model with $N=7$ and $S=10^4$
  (Fig. \ref{fig:choo_a}A). As a result we have
  \begin{equation}
    P_{\textrm{acc}} = \sum P_{\MIN}(\H) P(\beta, C(\H))
  \end{equation}
  \noindent where $P_{\MIN}(\H)$ is the probability of generating
  a scale with the average harmonicity score $\H$.
  By specifying a cost function $C(\H)$ and 
  $\beta$ we can then get the analytical probability
  distribution of $\H$ for the accepted scales. We use the Jensen-Shannon
  divergence, $\textrm{JSD}$, as a measure of selectivity. Using this method
  we study how the selectivity and acceptance rate change depending
  on the functional form of the bias.

  There are two basic ways of defining such a cost function for the \HAR~model
  such that as $\H$ increases the cost decreases:
  \begin{equation}
    C_1 = 1 - \H/A~,
  \end{equation}
  \noindent and 
  \begin{equation}
    C_2 = 1 / (A + \H)~
  \end{equation}
  \noindent where $A$ is a constant. Due to the form of the Boltzmann distribution
  if $C\leq0$ then $P=1$ regardless of $\beta$.
  This means that in general one should choose $A$
  such that $C>=0$ for all $\H$: for $C_1$ this is when $A \geq \H_{max}$;
  for $C_2$ this is when $A \geq - \H_{min}$. We can choose to violate this
  principle, with variable consequences. For $C_1$, the consequences are not
  so severe as the $\H$ distribution is positively skewed
  (Fig. \ref{fig:choo_a}A left) -- i.e., if $A$ is slightly 
  lower than $\H_{max}$ little will change because the probability
  of generating scales with $\H$ close to $\H_{max}$ is small,
  and the scales which end up with a negative cost are those
  with high $\H$ (the scales we wish to prioritize).
  This is because the form of $C_1$ acts to make scales with high $\H$
  have low cost.
  However for $C_2$, the consequences are severe (Fig. \ref{fig:choo_a}A right).
  In this case, the form of $C_2$ acts to penalize scales with low $\H$, i.e., 
  $C_2\to 0$ only as $\H \to\infty$, which is impossible as $\H$ is
  bounded by $\H_{max}$.
  If $A \leq - \H_{min}$ then scales with low $\H$ have $P=1$
  due to the change of sign in $C_2$.

  If we vary $A$ in the other direction ($A>\H_{max}$ for $C_1$, $A>-\H_{min}$ for $C_2$),
  we get a lower acceptance rate for a given selectivity (Fig. \ref{fig:choo_a}B-C).
  For $C_1$ changing $A$ in this way does not qualitatively alter
  the selectivity because the change is linear (Fig. \ref{fig:choo_a}A left -- the orange
  and purple lines are the same).
  However for $C_2$ changing $A$ in this way does qualitatively alter the
  selectivity (the orange and purple lines are not the same in Fig. \ref{fig:choo_a}A right).
  Thus, each cost function has a clear optimum value of $A$: at $A = \H_{max}$ for $C_1$, and
  at $A= -\H_{min}$ for $C_2$.

\begin{figure}
\centering
\includegraphics[width=\textwidth]{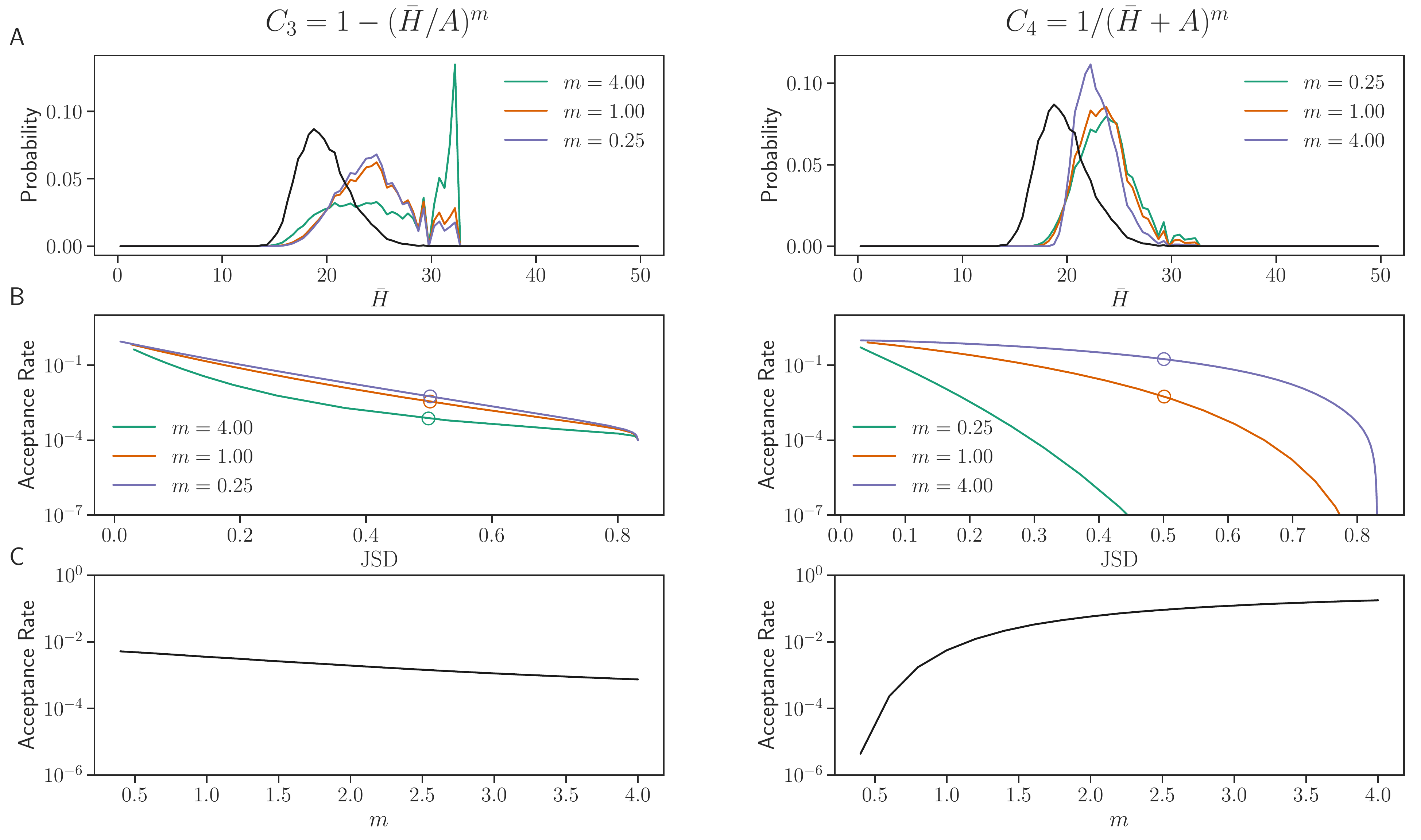}
\caption{\label{fig:choo_m}
  Effect of changing $m$ in the cost functions $C_3$ (left) and $C_4$ (right).
  A: Probability distribution of $\H$ for a population of $N=7$ scales
  accepted by the \MIN~model (black line). Theoretical probability
  distributions of $\H$ accepted by models with the cost functions
  $C_3$ (left) and $C_4$ (right). Values of $m$ correspond to:
  $m$ chosen so that the acceptance rate increases (green line),
  $m=1$ (orange line),
  $m$ chosen so that the acceptance rate decreases (purple line).
  $\beta$ is chosen so that the selectivity (JSD) is $0.5$.
  B: Acceptance rate vs. selectivity (JSD) for the same values of $m$ as in A.
  Circles correspond to a selectivity (JSD) of $0.5$.
  C: Dependence of the acceptance rate on $m$. Selectivity is constant at $0.5$.
}
\end{figure}

\begin{figure}
\centering
\includegraphics[width=\textwidth]{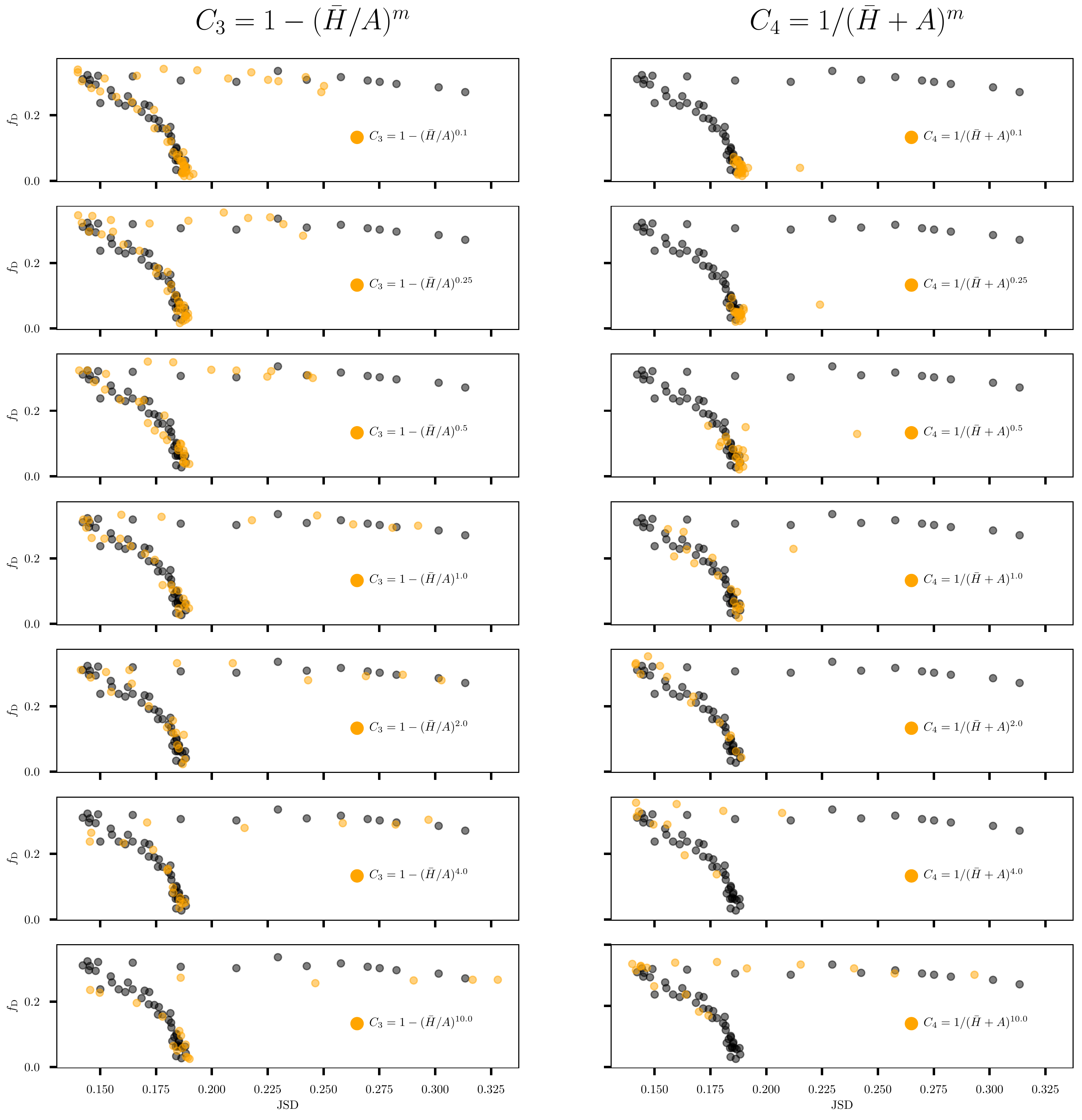}
\caption{\label{fig:bias_sens}
  Effect of changing $m$ on model performance for the cost functions $C_3$ (left) and $C_4$ (right),
  for $m = 0.1, 0.25, 0.5, 1, 2, 4, 10$. For each subplot we report the model
  performance as the Jensen-Shannon divergence between adjacent interval distributions, $\textrm{JSD}$,
  and the fraction of scales found by the model $\fD$. The black circles mark the
  results of the main \HAR~model. Increasing $\beta$ results in $\fD$
  increasing while $\dI$ decreases until the point of optimal results (top-left corner of each plot), after
  which $\dI$ sharply increases. After the optimal point, the fact that $\dI$ decreases while $\fD$
  is relatively constant indicates that when $\beta$ is higher
  than the optimum, the cost function is specialized for a subset of scales.
}
\end{figure}

  We can further optimize the cost function with an additional parameter $m$:
  \begin{equation}
    C_3 = 1 - (\H/A)^m~,
  \end{equation}
  \noindent and 
  \begin{equation}
    C_4 = 1 / (A + \H)^m~.
  \end{equation} 
  \noindent Changing $m$ has opposing effects for $C_3$ and $C_4$
  (Fig. \ref{fig:choo_m}B-C). For $C_3$, decreasing $m$ results in a converging, increasing acceptance
  rate, while increasing $m$ results in a diverging, decreasing acceptance rate.
  For $C_4$, increasing $m$ results in a converging, increasing acceptance rate,
  while decreasing $m$ results in a diverging, decreasing acceptance rate.
  Crucially, any changes to $m$ will result in different $\H$ distributions
  for a fixed selectivity. For all the theoretical $\H$ distributions
  in Fig. \ref{fig:choo_m}A the a selectivity is set at $0.5$, but they
  differ considerably. This should result in different results in
  our simulations. We investigated the effect of $m$ for $C_3$ and $C_4$
  on the performance of the corresponding models (Fig. \ref{fig:bias_sens}).
  We find that changing $m$ does change the performance of the models under
  some conditions. For $C_3$ ($C_4$) the
  results improve with decreasing (increasing) $m$, however this effect appears to
  converge at the level of performance of the \HAR~model presented in the main text.

\begin{table}\centering
\caption{\label{tab:fits}
  Goodness-of-fit statistics for how well the model distributions 
  fit the \DAT~distributions in the main text Fig. 3.
  For each model and each $N$ we show: sample size $S$ for \DAT~scales of size $N$;
  Jensen-Shannon divergence, $\textrm{JSD}$, between \DAT~ and model $\IP$ distributions
  (mean and $95\%$ confidence intervals); two sample Cram\'er-von Mises criterion (CvM) for
  each set of $\IP$ distributions (mean and $95\%$ confidence intervals);
  fraction of real scales found, $\fD$, (mean and $95\%$ confidence intervals).
}
\begin{tabular}{lllcccccc}
\toprule
Model  &  $N$  &  $S$  &  JSD  &  CI  &  CvM  &  CI  &  $f_D$  &  CI  \\ 
RAN    &   4 &  10 & 0.33 & (0.26, 0.40) & 0.74 & (0.47, 1.03) & 0.49 & (0.20, 0.80) \\
       &   5 & 180 & 0.32 & (0.30, 0.34) & 14.63 & (12.21, 17.42) & 0.07 & (0.03, 0.11) \\
       &   6 &  36 & 0.34 & (0.28, 0.41) & 3.60 & (2.56, 4.84) & 0.00 & (0.00, 0.00) \\
       &   7 & 480 & 0.35 & (0.34, 0.36) & 48.65 & (44.60, 53.07) & 0.01 & (0.00, 0.02) \\
       &   8 &  29 & 0.30 & (0.26, 0.34) & 3.24 & (2.75, 3.78) & 0.00 & (0.00, 0.00) \\
       &   9 &   7 & 0.30 & (0.27, 0.34) & 1.23 & (0.80, 1.79) & 0.00 & (0.00, 0.00) \vspace{0.1cm} \\
       & arithmatic mean &    & 0.34 & (0.33, 0.35) & 35.34 & (32.63, 38.36) & 0.03 & (0.02, 0.04) \vspace{0.2cm} \\
MIN    &   4 &  10 & 0.23 & (0.15, 0.31) & 0.44 & (0.22, 0.69) & 0.49 & (0.20, 0.80) \\
       &   5 & 180 & 0.19 & (0.17, 0.23) & 6.30 & (4.37, 8.69) & 0.27 & (0.21, 0.34) \\
       &   6 &  36 & 0.25 & (0.17, 0.32) & 1.68 & (0.95, 2.53) & 0.08 & (0.00, 0.17) \\
       &   7 & 480 & 0.19 & (0.18, 0.20) & 10.93 & (9.26, 12.99) & 0.02 & (0.01, 0.04) \\
       &   8 &  29 & 0.21 & (0.18, 0.25) & 2.55 & (1.46, 3.84) & 0.00 & (0.00, 0.00) \\
       &   9 &   7 & 0.31 & (0.28, 0.33) & 1.66 & (0.87, 2.30) & 0.00 & (0.00, 0.00) \vspace{0.1cm} \\
       & arithmatic mean &    & 0.20 & (0.19, 0.21) & 8.81 & (7.53, 10.24) & 0.09 & (0.07, 0.11) \vspace{0.2cm} \\
HAR    &   4 &  10 & 0.22 & (0.15, 0.31) & 0.40 & (0.20, 0.68) & 0.51 & (0.20, 0.80) \\
       &   5 & 180 & 0.16 & (0.13, 0.19) & 4.69 & (2.87, 7.04) & 0.51 & (0.44, 0.58) \\
       &   6 &  36 & 0.19 & (0.16, 0.25) & 0.69 & (0.36, 1.27) & 0.25 & (0.11, 0.42) \\
       &   7 & 480 & 0.14 & (0.13, 0.15) & 6.27 & (4.94, 7.90) & 0.31 & (0.27, 0.35) \\
       &   8 &  29 & 0.20 & (0.17, 0.22) & 2.60 & (1.44, 4.01) & 0.31 & (0.17, 0.48) \\
       &   9 &   7 & 0.26 & (0.25, 0.28) & 1.44 & (0.59, 2.21) & 0.57 & (0.14, 0.86) \vspace{0.1cm} \\
       & arithmatic mean &    & 0.15 & (0.14, 0.17) & 5.34 & (4.34, 6.60) & 0.36 & (0.33, 0.39) \vspace{0.2cm} \\
TRANS  &   4 &  10 & 0.25 & (0.18, 0.34) & 0.49 & (0.27, 0.76) & 0.40 & (0.10, 0.70) \\
       &   5 & 180 & 0.20 & (0.17, 0.23) & 6.43 & (4.42, 8.67) & 0.47 & (0.40, 0.55) \\
       &   6 &  36 & 0.20 & (0.15, 0.27) & 1.07 & (0.57, 1.75) & 0.33 & (0.19, 0.50) \\
       &   7 & 480 & 0.18 & (0.17, 0.19) & 9.04 & (7.19, 11.07) & 0.28 & (0.24, 0.32) \\
       &   8 &  29 & 0.22 & (0.18, 0.24) & 2.41 & (1.38, 3.61) & 0.48 & (0.31, 0.66) \\
       &   9 &   7 & 0.29 & (0.27, 0.30) & 1.52 & (0.81, 2.14) & 0.72 & (0.43, 1.00) \vspace{0.1cm} \\
       & arithmatic mean &    & 0.19 & (0.18, 0.20) & 7.57 & (6.28, 8.95) & 0.34 & (0.31, 0.37) \vspace{0.2cm} \\
FIF    &   4 &  10 & 0.25 & (0.17, 0.33) & 0.51 & (0.28, 0.77) & 0.49 & (0.20, 0.80) \\
       &   5 & 180 & 0.13 & (0.10, 0.16) & 2.41 & (1.26, 4.01) & 0.60 & (0.53, 0.67) \\
       &   6 &  36 & 0.20 & (0.14, 0.26) & 1.40 & (0.78, 2.16) & 0.30 & (0.17, 0.47) \\
       &   7 & 480 & 0.13 & (0.11, 0.14) & 5.82 & (4.78, 7.14) & 0.37 & (0.33, 0.41) \\
       &   8 &  29 & 0.17 & (0.14, 0.20) & 1.57 & (0.70, 2.69) & 0.45 & (0.28, 0.62) \\
       &   9 &   7 & 0.23 & (0.22, 0.25) & 0.98 & (0.40, 1.62) & 0.72 & (0.29, 1.00) \vspace{0.1cm} \\
       & arithmatic mean &    & 0.13 & (0.13, 0.15) & 4.49 & (3.72, 5.41) & 0.43 & (0.40, 0.46) \vspace{0.2cm} \\
\bottomrule
\end{tabular}
\end{table}

\clearpage

\bibliographystyle{unsrtnat}
\bibliography{database}